\documentclass[twocolumn,aps,prc,tightenlines,floats,floatfix,nofootinbib]{revtex4}%%

\def\bea{\begin{eqnarray}}
\def\eea{\end{eqnarray}}

\def\sfrac#1#2{{\textstyle \frac{#1}{#2}}}

\newcommand{\bra}[1]{\langle #1|}
\newcommand{\ket}[1]{|#1\rangle}

\def\be{\begin{equation}}
\def\ee{\end{equation}}
\def\ba{\begin{eqnarray}}
\def\ea{\end{eqnarray}}

\usepackage{graphics}
\usepackage{graphicx}
\usepackage{epsf}
\usepackage{amsmath}
\usepackage{amssymb}
\begin{document}

\phantom{0}
\vspace{-0.2in}
\hspace{5.5in}
\parbox{1.5in}{ \leftline{JLAB-THY-10-1134}}

\vspace{-1in}%\parbox{1.5in}{ \vspace{-9.6in}}  % moves the preprint box down

\title
{\bf Valence quark contributions for the
$\gamma N \to P_{11}(1440)$ form factors }

\author{
G.~Ramalho$^1$ and K.~Tsushima$^2$
\vspace{-0.1in}  }

\affiliation{
$^1$Centro de F{\'\i}sica Te\'orica e de Part{\'\i}culas,
Av.\ Rovisco Pais, 1049-001 Lisboa, Portugal \vspace{-0.15in} }

\affiliation{
$^2$Excited Baryon Analysis Center (EBAC) and Theory Center,
Thomas Jefferson National Accelerator Facility, Newport News,
VA 23606, USA}

\vspace{0.2in}
\date{\today}

\phantom{0}

\begin{abstract}
Using a covariant spectator quark model we estimate
valence quark contributions to the $F_1^\ast(Q^2)$ and $F_2^\ast(Q^2)$
transition form factors for the $\gamma N \to P_{11}(1440)$ reaction.
The Roper resonance, $P_{11}(1440)$, is assumed to be
the first radial excitation of the nucleon.
The present model requires no extra parameters
except for those already fixed by previous studies of the nucleon.
Our results are consistent with the experimental data in the
high $Q^2$ region, and those from lattice QCD.
We also estimate the meson cloud contributions,
focusing on the low $Q^2$ region, where they are
expected to be dominant.

\end{abstract}

%\phantom{0}
%\vspace{7.0in}
%\vspace{-6in}
\vspace*{0.9in}  % sets how far the title is below the preprint box
\maketitle

\section{Introduction}

The structure of the $P_{11}(1440)$ resonance, which will be referred to as "the Roper" in this article,
has been a long-standing mystery.
Its mass and large width are difficult to understood
in a framework of quark models~\cite{Clement06,Sarantsev08,Suzuki09,Aznauryan07}.
The decay branches imply that it is a mixture of
$\pi N$ and $\pi \pi N$ states
(dominated by the mixture of $\pi \Delta$ and $\sigma N$ channels).
Recently, the precision data from CLAS at JLab~\cite{CLAS}
(single and double pion electroproduction),
and a global analysis from MAID~\cite{MAID,Tiator09}
in the region $Q^2 \le 6$ GeV$^2$ have become available.
The extracted $\gamma N \to P_{11}(1440)$ helicity amplitudes
can give hints on the internal structure of the Roper.
Indeed, the amplitudes (and the transition form factors) extracted from the experiments,
turned out to show a surprising $Q^2$ dependence.

In a simple quark model picture, the Roper
can be regarded as the first radial excitation of the nucleon.
Recent experimental data at large $Q^2$~\cite{CLAS,MAID}
support this picture~\cite{CLAS,Aznauryan07,Aznauryan08,Burkert08}.
In the high $Q^2$ region ($Q^2 > 2$ GeV$^2$) the data
is consistent with the predicted $Q^2$ dependence of
relativistic constituent and
light-front quark models~\cite{CLAS,Aznauryan07,Aznauryan08}.
These models, however, fail to explain the lower $Q^2$ data,
particularly at the photon point, $Q^2=0$.
This discrepancy has been interpreted as
a manifestation of the missing meson cloud effects
in the models~\cite{Aznauryan07,Aznauryan08,Burkert08,JDiaz08}.
Unfortunately, the magnitude of the meson cloud contributions
is less constrained,
and they are more model dependent than the
valence quark contributions, which
are dominant at larger $Q^2$.
(In general, the meson cloud contributions are expected to
fall as $Q^2$ increases.)
In this study, we estimate the meson cloud contributions
using a covariant spectator constituent quark model
which is consistent with the large $Q^2$ region data of the CLAS.
The present model requires no extra parameters
except for those already fixed by the previous
studies for the nucleon~\cite{Nucleon}.

The $\gamma N \to P_{11}(1440)$ transition
amplitudes have been studied in several ways: treating the Roper
as a quark-gluon system~\cite{Li92}, a
pure valence quark system~\cite{Warns90,Weber90,Capstick95,Cardarelli98,Giannini01,JDiaz04,Aznauryan07},
and a valence quark system dressed by $q \bar q$ pairs~\cite{Cano98,Li06,Santopinto08,Dillig04,JDiaz06}.
Relativity has proven to be very important,
not only for a large $Q^2$, but also for the region $Q^2 \sim 0$.
It may even decide the sign of the amplitudes near
$Q^2=0$~\cite{Warns90,Capstick95,CLAS,Aznauryan07}.
The nucleon to Roper transition amplitudes have been also studied using
dynamical baryon-meson coupled-channel models.
%%%%%%\cite{Burkert04,Krehl00,JDiaz07b,JDiaz08,Golli08a,Julich,Kamalov01}
They usually treat the baryons and mesons as effective degrees of freedom, and the baryons
are dressed with a meson cloud nonperturbatively.
(For details, see e.g., Refs.~\cite{JDiaz08,Krehl00,JDiaz07b,JDiaz07,JDiaz09,Suzuki09b,Golli08a,Golli09,Julich,Kamalov01} and
a review~\cite{Burkert04}.)
The nonvalence quark degrees of freedom was also
introduced by coupling the pion
and other meson fields to the valence quarks~\cite{Bermuth90,Dong99,Chen08}.
The $\gamma N \to P_{11}(1440)$ transition was also studied
in lattice QCD~\cite{Lin09}.

In this study we compute the
$\gamma N \to P_{11}(1440)$ transition form factors using a
valence quark model based on the covariant
spectator formalism~\cite{Gross}.
Relativity is implemented consistently.
In this model a baryon is described as a quark-diquark system,
where the diquark has all possible polarizations (spin state 0 or 1)
and acts as a spectator. The isolated quark in the baryon interacts with the photon
in the impulse approximation.
The model has been applied successfully for the studies of
spin 1/2 and spin 3/2 low-lying baryons~\cite{Nucleon,NDelta,NDeltaD,DeltaFF0,DeltaFF,Omega,Octet}.
We assume that the Roper is the first radial excitation
of the nucleon, which also ensures that it is orthogonal to the
nucleon (valence quark) state.
Using a model with no extra new adjustable parameters,
we compute the $Q^2$ dependence of
the transition form factors.
We find an excellent agreement with the data~\cite{CLAS,MAID}
in the high $Q^2$ region.
Furthermore, we extend the model to
a lattice QCD regime and compare with the lattice data
for a large pion mass ($m_\pi = 720$ MeV),
and find also a good agreement,
particularly in the low $Q^2$ region, $Q^2 < 1$ GeV$^2$.
Note that under the two conditions,
high $Q^2$ and lattice simulations with heavy pions,
meson cloud effects are expected to become small or negligible~\cite{Wang:2007iw}.
The two agreements mentioned above, support that the present model can
describe well the valence quark contributions.
Encouraged by the successful features of the model,
we estimate the meson cloud contributions focusing on the low $Q^2$ region.

This article is organized as follows:
In Sec.~\ref{secAmp} general relations among the current,
helicity amplitudes, and transition form factors are given.
Formalisms, wave functions, and explicit expressions for
the transition form factors are presented in Sec.~\ref{secFF}.
In Sec.~\ref{secResults} numerical results are presented, and a comparison is made
with the lattice simulation data.
Furthermore, meson cloud contributions are estimated in the present approach.
Finally, discussions and conclusions are given in Sec.~\ref{secConclusions}.

%\newpage

\section{Helicity amplitudes and form factors}
\label{secAmp}

The electromagnetic transition between a nucleon (mass $M$)
and a spin 1/2 positive parity nucleon resonance $N^\ast$ (mass $M_R$)
can be written using the Dirac $F_1^\ast$ and Pauli $F_2^\ast$ type
form factors, defined by the current~\cite{Tiator09},
\ba
& &
\hspace{-0.5cm}
J^\mu=
\bar u_R (P_+) \left\{
\left( \gamma^\mu - \frac{\not \! q q^\mu}{q^2}\right)F_1^\ast%(Q^2)
+
\frac{i \sigma^{\mu \nu} q_\nu}{M_R+ M} F_2^\ast%(Q^2)
\right\} u(P_-),
\nonumber \\
& &
\hspace{-0.5cm}
\label{eqJgen}
\ea
where, $u_R$ ($u$) is the
$N^\ast$ (nucleon) Dirac spinor,
$P_+$ ($P_-$) is the final (initial) momentum
and $q=P_+-P_-$.
Spin projection indices are suppressed for simplicity.

Usually, experimental data measured for hadron electromagnetic
structure are reported in terms of
the helicity amplitudes in a particular frame.
The most popular choice is
the rest frame of the final state, projecting
the current on the photon polarization states,
$\varepsilon_\lambda^\mu$,
where $\lambda=0,\pm$ is the photon spin projection.
In the $N^\ast$ rest frame, the helicity amplitudes, $A_{1/2}$ and
$S_{1/2}$, are given by~\cite{Aznauryan07,Aznauryan08,Tiator09}:
\ba
& &
A_{1/2}(Q^2)= {\cal K} \frac{1}{e}
\bra{N^\ast,+\sfrac{1}{2}} \varepsilon_+ \cdot J
\ket{N, -\sfrac{1}{2}},
\label{eqA1}\\
& &
S_{1/2}(Q^2)= {\cal K} \frac{1}{e}
\bra{N^\ast,+\sfrac{1}{2}} \varepsilon_0 \cdot J
\ket{N, + \sfrac{1}{2}} \frac{|{\bf q}|}{Q}.
\label{eqS0}
\ea
Here, $e=\sqrt{4\pi \alpha}$ is the magnitude of the
electron charge with $\alpha \simeq 1/137$,
and
\be
{\cal K}= \sqrt{\frac{2\pi \alpha}{K}},
\ee
with $K=\frac{M_R^2-M^2}{2M_R}$. $|{\bf q}|$ is
the photon momentum in the $N^\ast$ rest frame,
\be
|{\bf q}|= \frac{\sqrt{Q_+^2Q_-^2}}{2M_R},
\label{eqq2}
\ee
where $Q_\pm^2= (M_R \pm M)^2 + Q^2$, with $Q^2=-q^2$.

The helicity amplitudes, $A_{1/2}$ and $S_{1/2}$,
can be related with the form factors $F_1^\ast$
and $F_2^\ast$ via
Eqs.~(\ref{eqJgen})-(\ref{eqS0})~\cite{Aznauryan07,Aznauryan08,Tiator09}:
\ba
& &
\hspace{-1.2cm}
A_{1/2}(Q^2)=
{\cal R} \left\{ F_1^\ast (Q^2) + F_2^\ast(Q^2) \right\},
\label{eqA12}\\
& &
\hspace{-1.2cm}
S_{1/2}(Q^2)=
\frac{ {\cal R} }{\sqrt{2}}
|{\bf q}| \frac{M_R+M}{Q^2}
\left\{ F_1^\ast (Q^2) -\tau F_2^\ast(Q^2) \right\},
\label{eqS12}
\ea
where $\tau=\sfrac{Q^2}{(M_R+M)^2}$, and
\ba
{\cal R}= \sqrt{\frac{\pi \alpha Q_-^2}{M_R M K}}.
\ea
Note that the amplitude $S_{1/2}$ is
determined by virtual photons and not specified
at $Q^2=0$, but obtained only in the limit $Q^2 \to 0$.
In this case, one has
\ba
& & \hspace{-0.7cm}
A_{1/2}(0) = {\cal R} F_2^\ast(0),
\label{eqA0} \\
& &  \hspace{-0.7cm}
S_{1/2}(0) = \frac{{\cal R}}{\sqrt{2}} (M_R+M) K
\left\{
\frac{d F_1^\ast}{d Q^2}(0) - \frac{F_2^\ast(0)}{(M_R+M)^2}
\right\}. \nonumber \\
& &
\ea

The inverse relations
for $F_i^\ast (i=1,2)$ in terms of the helicity amplitudes are:
\ba
& &
\hspace{-0.4cm}
F_1^\ast(Q^2) =\frac{1}{\cal R} \frac{\tau}{1+\tau}
\left\{
A_{1/2}(Q^2) + \sqrt{2}\frac{M_R+M}{|{\bf q}|} S_{1/2}(Q^2)
\right\}, \nonumber \\
& &
\hspace{-0.4cm}
\label{eqF1}  \\
& &
\hspace{-0.4cm}
F_2^\ast (Q^2)=\frac{1}{\cal R} \frac{1}{1+\tau}
\left\{
A_{1/2} (Q^2) - \sqrt{2}\frac{M_R+M}{|{\bf q}|} \tau S_{1/2} (Q^2)
\right\}. \nonumber \\
& &
\hspace{-0.5cm}
\label{eqF2}
\ea
If the amplitudes $A_{1/2}$ and $S_{1/2}$ are finite
at $Q^2=0$, Eq.~(\ref{eqF1}) implies that
\be
F_1^\ast(0)=0.
\ee
As for $F_2^\ast (0)$, there is no particular condition.

For the Roper a fit to the data suggests a
small value for $S_{1/2}(0)$~\cite{Tiator09}.
Thus, at low $Q^2$,  $F_1^\ast$ and  $F_2^\ast$
are determined essentially by $A_{1/2}(Q^2)$.

\section{Form factors in a valence quark model}
\label{secFF}

So far, a covariant spectator quark model has been developed and
applied successfully to the spin 1/2 \cite{Nucleon,Octet}
and 3/2~\cite{NDelta,NDeltaD,DeltaFF0,DeltaFF,Omega} ground states
with no radial excitations.
In the model, a three-quark baryon is
described as an isolated quark which interacts with a photon, and a spectator diquark.
The wave function of a baryon is represented in terms
of the flavor and spin states of the quark and diquark
combined with the relative angular momentum.
In the spirit of the covariant spectator theory
the quark pair degrees of freedom are integrated out
to form an on-shell diquark with a certain mass $m_D$.

\subsection{Baryon wave functions}

To describe the momentum distribution
of the quark-diquark system in a baryon $B$,
we introduce a scalar wave function $\psi_B$,
which depends on the relative angular
momentum and the radial excitation of the system.
As the baryon and the diquark are
on-shell the scalar wave function $\psi_B$
can be written as a function of $(P-k)^2$
\cite{Nucleon}, where $P$ is
the baryon total momentum and $k$ the diquark momentum.
We represent that dependence
in term of the
dimensionless variable \cite{Nucleon},
\be
\chi_B= \frac{(M_B-m_D)^2-(P-k)^2}{M_B m_D},
\label{eqChi}
\ee
where $B$ is the baryon index ($B=N,\Delta, N^\ast$, etc)
and $M_B$ the mass.

The nucleon wave function, $\Psi_N(P,k)$,
can be written in the simplest S-wave model~\cite{Nucleon}:
\be
\Psi_N(P,k)=
\frac{1}{\sqrt{2}}
\left[
\phi_I^0 \phi_S^0 +
\phi_I^1 \phi_S^1
\right] \psi_N(P,k),
\label{eqPSIN}
\ee
where $\phi^{0,1}_{I,S}$ represents isospin ($I$)
or spin ($S$) states corresponding to the
total magnitude of either 0 or 1 in the diquark configuration~\cite{Nucleon}.
(See Appendix~\ref{appSQM} for the detail.)
The wave function represented by Eq.~(\ref{eqPSIN}),
satisfies the Dirac equation~\cite{Nucleon,NDelta}.
The scalar wave function $\psi_N$ (S state)  is given by
\be
\psi_N(P,k) = \frac{N_0}{m_D(\beta_1 + \chi_N)(\beta_2 +\chi_N)},
\label{eqPsiN}
\ee
where $\chi_N$ is obtained by inserting $M_B=M$ in Eq.~(\ref{eqChi}),
and $N_0$ the normalization constant.
In a parameterization where $\beta_1 > \beta_2$,
$\beta_1$ is associated with the long range
physics, while $\beta_2$ the short range physics.

In the present approach, the spin and isospin of the Roper
are the same as those of the nucleon.
Assuming the Roper to be the first radial excitation of the nucleon,
we can write the Roper scalar wave function $\psi_R$ as
\ba
\psi_R(P,k) &=&
N_R \frac{\beta_3 -\chi_R}{(\beta_1+ \chi_R)} \nonumber \\
& &
\times
\frac{1}{m_D(\beta_1 + \chi_R)(\beta_2 +\chi_R)},
\label{eqPsiR}
\ea
where $\beta_3$ is a new parameter which will be discussed later, and
$N_R$ the normalization constant.
The sign of the normalization constant can be fixed
by the experimental data~\cite{Aznauryan08}.
This particular form with the order one polynomial in  $\chi_R$ (as $\beta_3-\chi_R$),
is motivated by the harmonic-oscillator potential model
for the three-quark system.
A similar dependence was adopted in Ref.~\cite{JDiaz04}.

Note that the wave function represented by Eq.~(\ref{eqPsiR})
preserves the short range
behavior as presented in the nucleon scalar wave function,
and that it simultaneously modifies
the long range properties through the factor
$\sfrac{\beta_3 -\chi_R}{(\beta_1+\chi_R)}$.
The Roper wave function $\Psi_R$ also satisfies
the Dirac equation with mass $M_R$.

The parameter $\beta_3$ in
the Roper scalar wave function in Eq.~(\ref{eqPsiR}),
will be fixed by the orthogonality
condition between the nucleon and the Roper states.
$N_R$ will be fixed by $\int_k |\psi_R|^2=1$ at $Q^2=0$,
the same as that for the nucleon~\cite{Nucleon}.
Thus,
the description of the Roper
requires no extra parameters to be fixed for the present purpose,
since the parameters $\beta_{1}$ and $\beta_2$
have already been fixed~\cite{Nucleon}.

\begin{figure*}[t]
\vspace{.5cm}
\centerline{
\mbox{
\includegraphics[width=3.2in]{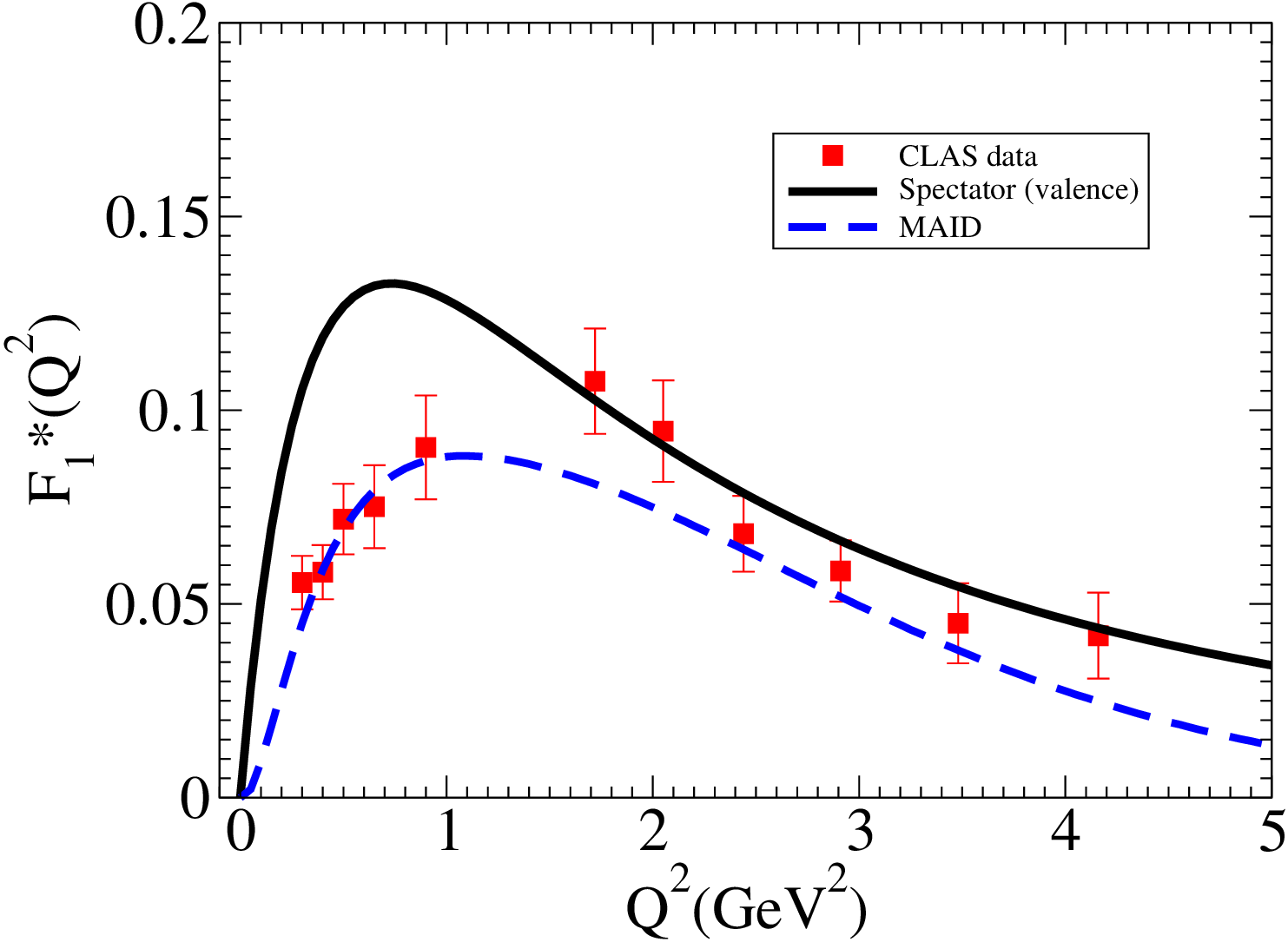}  \hspace{.1cm}
\includegraphics[width=3.2in]{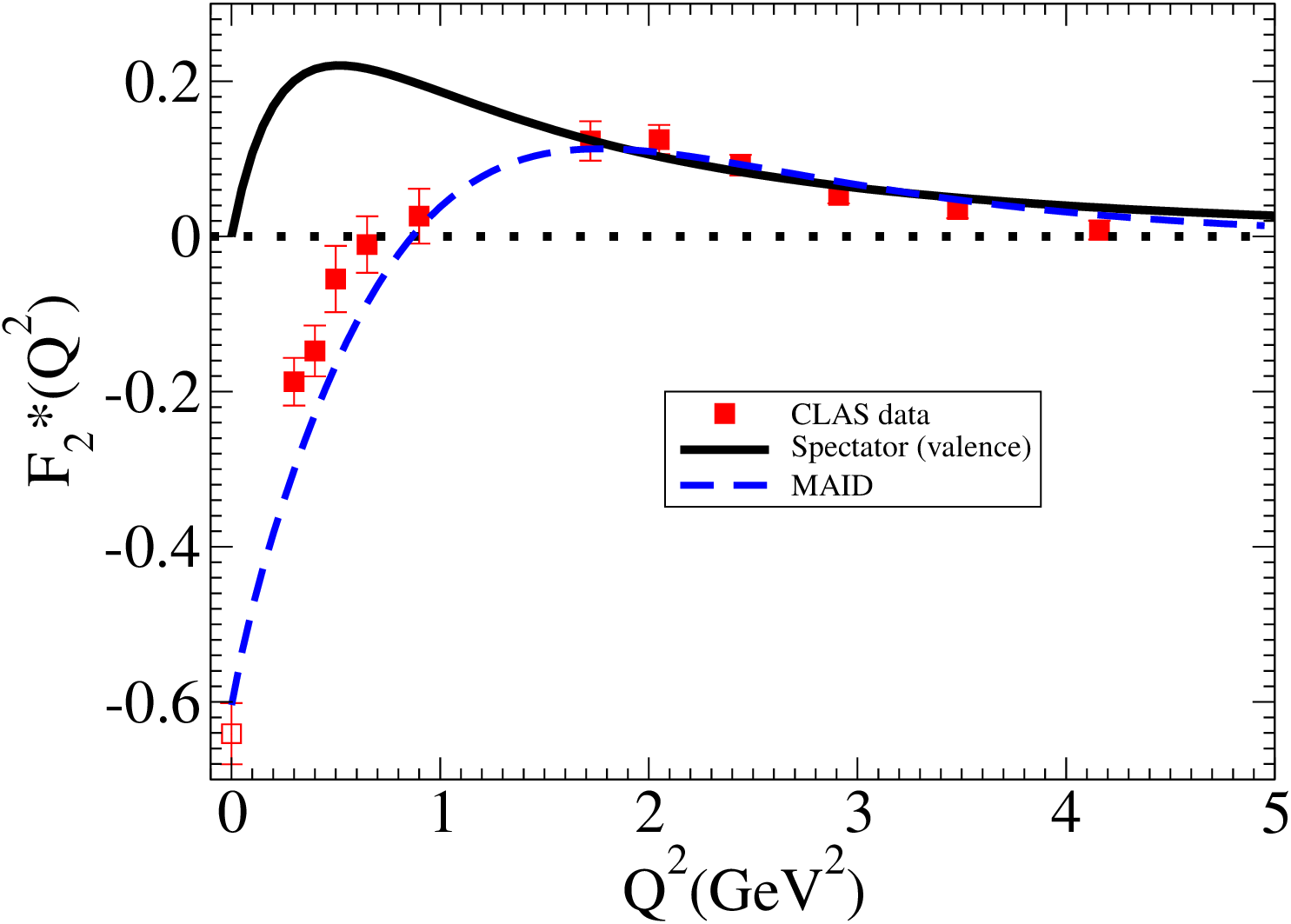}
}}
\caption{\footnotesize{
Valence quark contributions (solid lines) calculated in the present model
for $F_1^\ast$ and $F_2^\ast$.
The CLAS data~\cite{CLAS} (squares with error bars) and the MAID fit (dashed lines) are also shown.
The result for $F_2^\ast(0)$ is obtained
from the PDG result~\cite{PDG} by converting the $A_{1/2}(0)$.
[See Eq.~(\ref{eqA0}].)
}}
%\vspace{-1cm}
\label{figF1F2}
\end{figure*}

\begin{figure}[t]
\vspace{.5cm}
\centerline{
\mbox{
\includegraphics[width=3.2in]{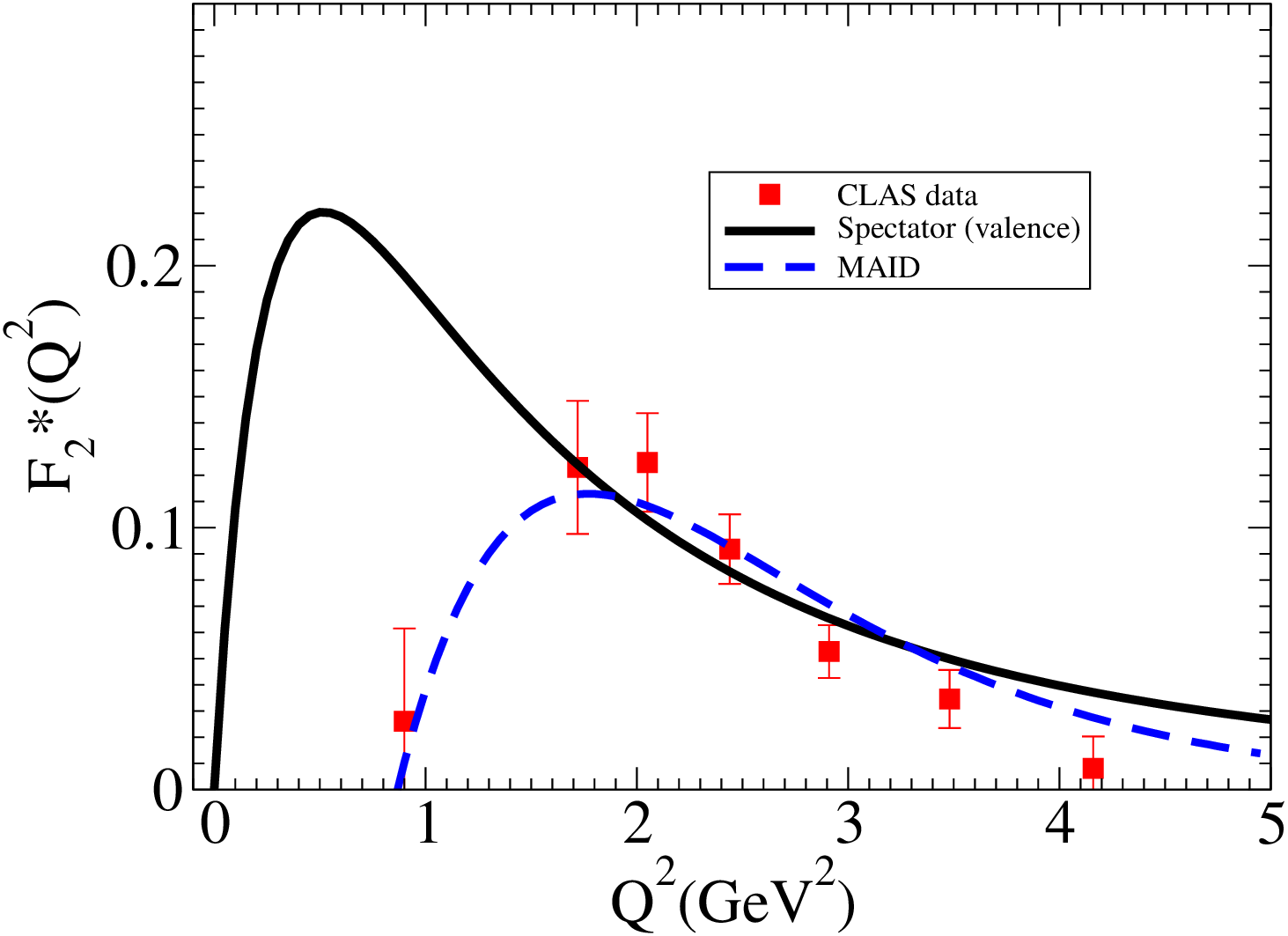}}}
\caption{\footnotesize{
Valence quark contributions for
$F_2^\ast$, but partly magnified from Fig.~\ref{figF1F2}.
See also the caption of Fig.~\ref{figF1F2}.
}}
%\vspace{-1cm}
\label{figF2}
\end{figure}

\subsection{Transition form factors}

\begin{figure*}[t]
\vspace{.5cm}
\centerline{
\mbox{
\includegraphics[width=3.2in]{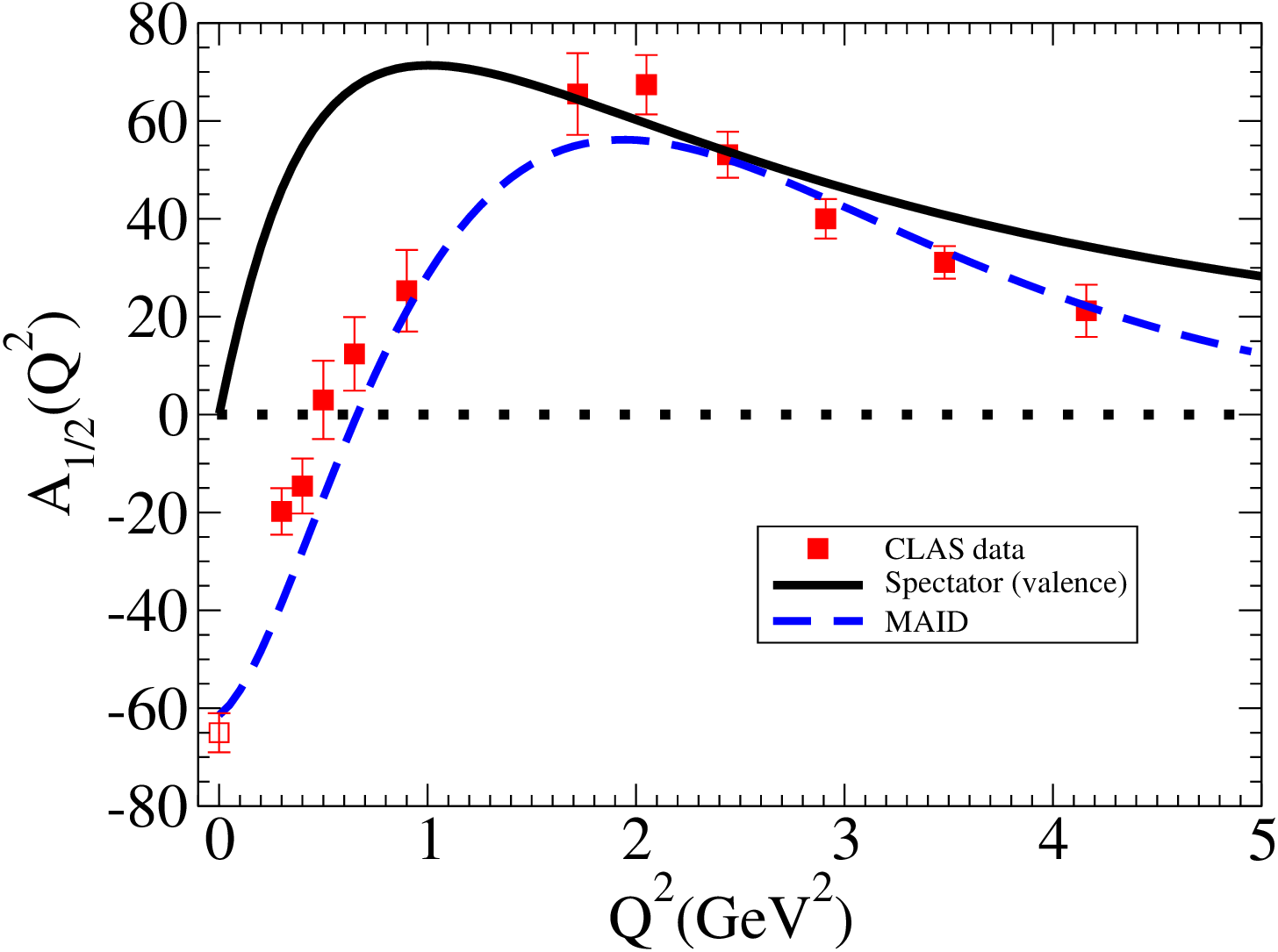}  \hspace{.1cm}
\includegraphics[width=3.2in]{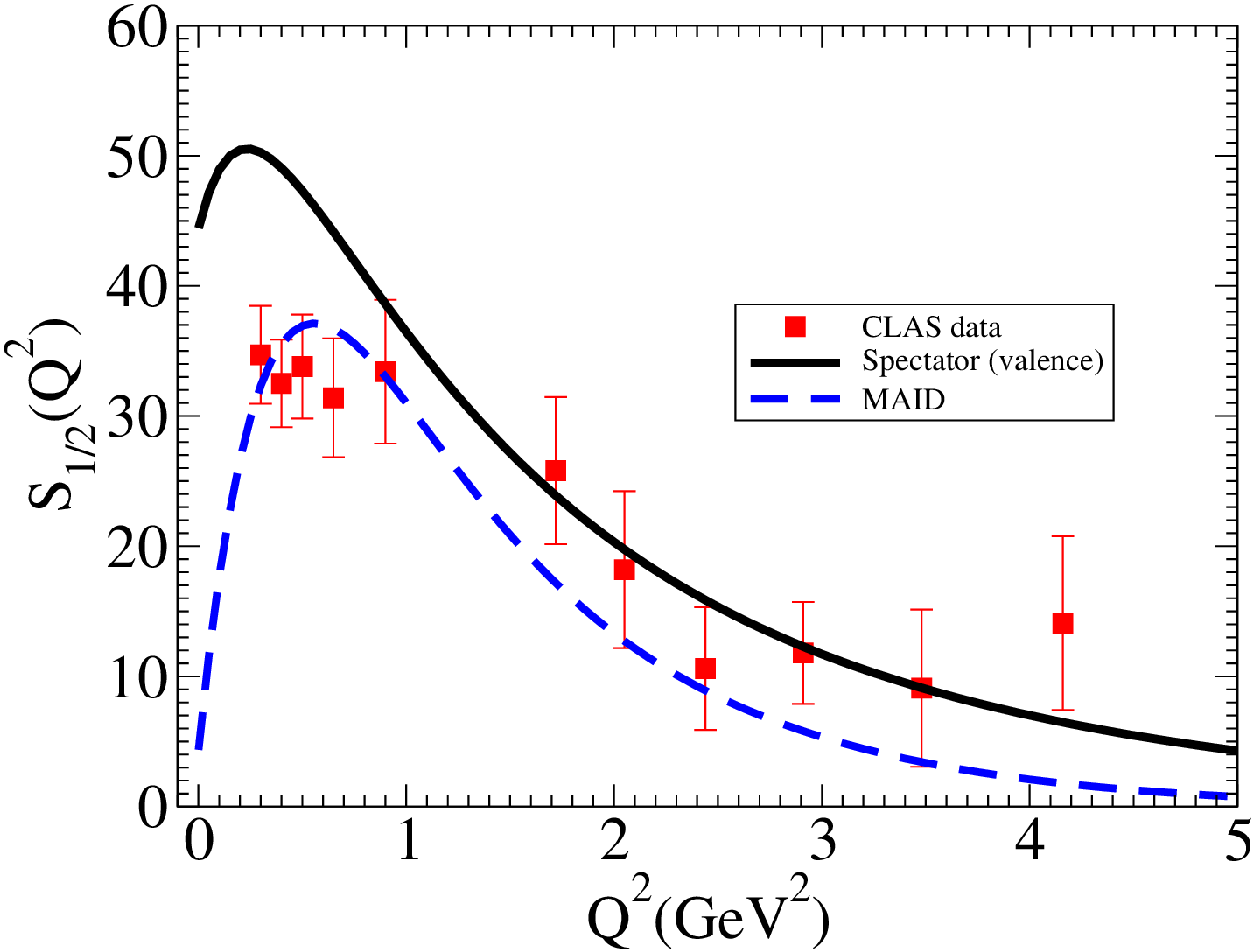}
}}
\caption{\footnotesize{
Valence quark contributions for the
$A_{1/2}$ and $S_{1/2}$ helicity amplitudes
in units of 10$^{-3}$ GeV$^{-1/2}$.
The CLAS data~\cite{CLAS} and the MAID fit are also shown~\cite{Tiator09}.
See also the caption of Fig.~\ref{figF1F2}.
}}
%\vspace{-1cm}
\label{figA12S12}
\end{figure*}

In the covariant spectator quark model
the transition current for the $\gamma N \to P_{11}(1440)$
reaction can be written in a
relativistic impulse approximation~\cite{Nucleon,NDelta,NDeltaD},
\be
J^\mu=
3 \sum_{\lambda} \int_k
%\bar \Psi_{R}(P_+,k)
\overline \Psi_{R}(P_+,k)
j_I^\mu \Psi_N(P_-,k),
\label{eqJsp}
\ee
where $\int_k \equiv \int \sfrac{d^3 k}{2E_D(2\pi)^3}$
with $E_D$ the diquark on-shell energy, $P_+ - P_- = q$ ($Q^2=-q^2$),
and $j_I^\mu$ is the quark current
parameterized as
\be
j_I^\mu= j_1(Q^2) \gamma^\mu +
j_2(Q^2)  \frac{i \sigma^{\mu \nu} q_\nu}{2M}.
\label{eqjI}
\ee
The Dirac ($j_1$) and Pauli ($j_2$) quark form factors in the above
are also decomposed as
\be
j_i(Q^2)= \frac{1}{6}f_{i+}(Q^2) +   \frac{1}{2}f_{i-}(Q^2) \tau_3,
 \hspace{1em} (i=1,2).
\label{eqjI2}
\ee
The quark form factors $f_{i\pm}$
are normalized as $f_{1\pm}(0)=1$ and $f_{2\pm}(0)=\kappa_\pm$
(isoscalar and isovector quark anomalous moments).
Their  explicit expressions can be found in
Refs.~\cite{Nucleon,NDelta,NDeltaD}
and in Appendix~\ref{appSQM}.

Using the expressions
for the nucleon and Roper wave functions,
Eqs.~(\ref{eqPSIN})-(\ref{eqPsiR}),
and the hadronic current, Eq.~(\ref{eqJsp}),
one gets explicit expressions for $F_1^\ast (Q^2)$ and $F_2^\ast (Q^2)$:
\ba
F_1^\ast (Q^2)&=&
\frac{3}{2} j_1 {\cal I}
+ \frac{1}{2}\frac{3(M_R+M)^2-Q^2}{(M_R+M)^2+Q^2} j_3 {\cal I}
\nonumber \\
& &
 - \frac{M_R+M}{M} \frac{Q^2}{(M_R+M)^2+Q^2} j_4 {\cal I},
\label{eqF1b} \\
F_2^\ast (Q^2)&=&
\frac{3}{4} \frac{M_R+M}{M} j_2 {\cal I}
- \frac{(M_R+M)^2}{(M_R+M)^2+Q^2} j_3 {\cal I} \nonumber \\
& &
 + \frac{M_R+M}{2M} \frac{(M_R+M)^2-3 Q^2}{(M_R+M)^2+Q^2} j_4 {\cal I},
\label{eqF2b}
\ea
where ${\cal I}(Q^2)$ is
the overlap integral for the Roper and nucleon scalar wave functions:
\be
{\cal I}(Q^2)= \int_k \psi_R(P_+,k) \psi_N (P_-,k),
\ee
and
\ba
j_{(i+2)}&=& \frac{1}{3}\tau_3 j_i\tau_3 \nonumber \\
       &=& \frac{1}{6}f_{i+}(Q^2) -   \frac{1}{6}f_{i-}(Q^2) \tau_3, \hspace{1em} (i=1,2).
\label{eqjII}
\ea

Equations~(\ref{eqF1b})-(\ref{eqF2b}) are
the main expressions of the present model
for the transition form factors.
%for the $\gamma N \to P_{11}(1440)$ reaction.
At $Q^2=0$, one has
\ba
F_1^\ast(0)&=& \frac{3}{2}(j_1+j_3) {\cal I} (0), \nonumber  \\
          &=& \frac{1}{2}(1+ \tau_3)  {\cal I}(0).
\ea
In this case, the desired result,
$F_1^\ast(0)=0$, is ensured only if ${\cal I}(0)=0$.
But this is indeed fulfilled by the
orthogonality condition between the nucleon and the Roper wave functions.
More detail will be discussed in next section.

% Figures 1 and 2

\subsection{Orthogonality condition between the nucleon and the Roper states}

The orthogonality condition between the nucleon and the Roper states
in the present approach is:
\be
{\cal I}(0)= \int_k \psi_R(\bar P_+,k) \psi_N(\bar P_-,k)=0.
\label{eqI0}
\ee
This can be regarded as a generalization of the nonrelativistic
orthogonality condition when $Q^2=0$
in the final Roper rest frame, namely, $\bar P_+=(M_R,0,0,0)$ and
$\bar P_-= \left(\sfrac{M_R^2 + M^2}{2M_R},0,0,-\sfrac{M_R^2-M^2}{2M_R}\right)$.
In Appendix~\ref{appOrth}, we discuss more on
Eq.~(\ref{eqI0}) and the
nonrelativistic orthogonality condition.
Note that the condition Eq.~(\ref{eqI0})
is not automatically verified for the nucleon and the
Roper wave functions in
Eqs.~(\ref{eqPsiN})-(\ref{eqPsiR}).
The orthogonality condition is satisfied
only for a particular value of $\beta_3$
that sets  ${\cal I}(0)=0$.

A direct consequence of the orthogonality condition Eq.~(\ref{eqI0}),
is that $F_1^\ast(0)=0$, and also that $F_2^\ast(0)=0$.
$F_1^\ast(0)=0$ is consistent with the general properties
of the $\gamma N \to P_{11}(1440)$ reaction.
$F_2^\ast(0)=0$ is a prediction of our model
as a consequence of the orthogonality condition Eq.~(\ref{eqI0}), but it also
corresponds to an approximation, since the sea quark contributions are ignored
in the present approach.
A more accurate treatment would give a small value for $|F_2^\ast(0)|$.

\section{Results}
\label{secResults}

The nucleon and the Roper wave functions are
described by Eq.~(\ref{eqPSIN})
with their scalar functions, Eqs.~(\ref{eqPsiN})-(\ref{eqPsiR}).
The parameters for the Roper wave function are determined by
those of the nucleon~\cite{Nucleon}:
$\beta_1=0.049$ and $\beta_2=0.717$,
and by the orthogonality condition Eq.~(\ref{eqI0}) to give $\beta_3=0.130$
and $N_R=3.35$.

% Figure 4

\subsection{Results for $F_1^\ast$ and $F_2^\ast$}
\label{secPhysical}

Before presenting the results, we emphasize again that
there are no new free parameters
to be fixed in the model.
Thus, for a given $Q^2$, the form factors can be calculated
using Eqs.~(\ref{eqF1b})-(\ref{eqF2b}).
The valence quark contributions calculated in the present model
for the $F_1^\ast$ and $F_2^\ast$ form factors are shown
in Fig.~\ref{figF1F2} by the solid lines.
In Fig.~\ref{figF2} also a part magnified from Fig.~\ref{figF1F2} for $F_2^\ast$
is shown.
The magnitudes of the present results are consistent
with constituent and light-front quark models~\cite{CLAS,Aznauryan07,JDiaz04}.
For convenience, we present in Fig.~\ref{figA12S12}
the helicity amplitudes calculated in the Roper rest frame,
using the relations Eqs.~(\ref{eqA12})-(\ref{eqS12}).

In Fig.~\ref{figF1F2} we also compare the
valence quark contributions with the CLAS data for $F_1^\ast$ and $F_2^\ast$.
The CLAS data were extracted combining dispersion relations
with a unitary isobar model analysis~\cite{CLAS}.
As one can see our result is very close to the data in
the region $Q^2>1.5$ GeV$^2$ (see especially Fig.~\ref{figF2} for $F_2^\ast$),
which supports the idea that the meson cloud
contributions are suppressed in the high $Q^2$ region, and the assumption that
the Roper is the first radial excitation of the nucleon.
This achievement may be impressive since we have introduced no extra
new parameters as already mentioned.
One only may identify the long range behavior
of the Roper wave function with that of the nucleon,
and may ensure the orthogonality
between the Roper and the nucleon wave functions.
In Fig.~\ref{figF1F2} we also show a fit to the
data from Ref.~\cite{MAID}, using the code MAID2007
(abbreviated MAID)~\cite{Tiator09}.
Both the CLAS data and the MAID fit are similar for $Q^2> 2$ GeV$^2$ region,
although some differences may be noticeable,
in particular for $F_1^\ast$.
This discrepancy in $F_1^\ast$ can be also a
consequence of the different data sets
used in the analysis between the CLAS data and the MAID fit.
Our predictions for $F_1^\ast$ are closer to the CLAS analysis than to
the MAID fit, although the differences
are comparable with the error bars of the CLAS data.

\begin{figure*}[t]
\vspace{.5cm}
\centerline{
\mbox{
\includegraphics[width=3.2in]{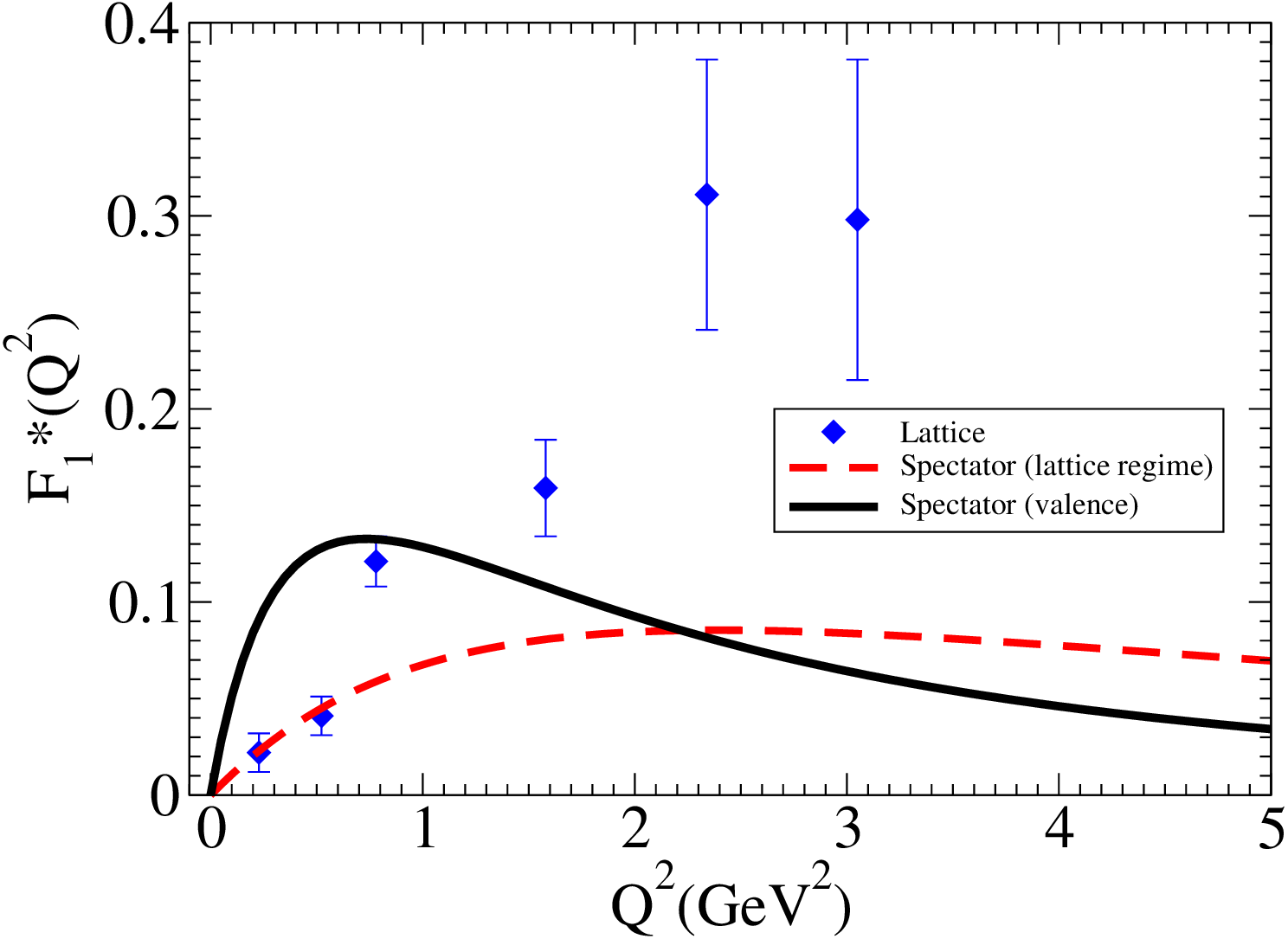}  \hspace{.1cm}
\includegraphics[width=3.2in]{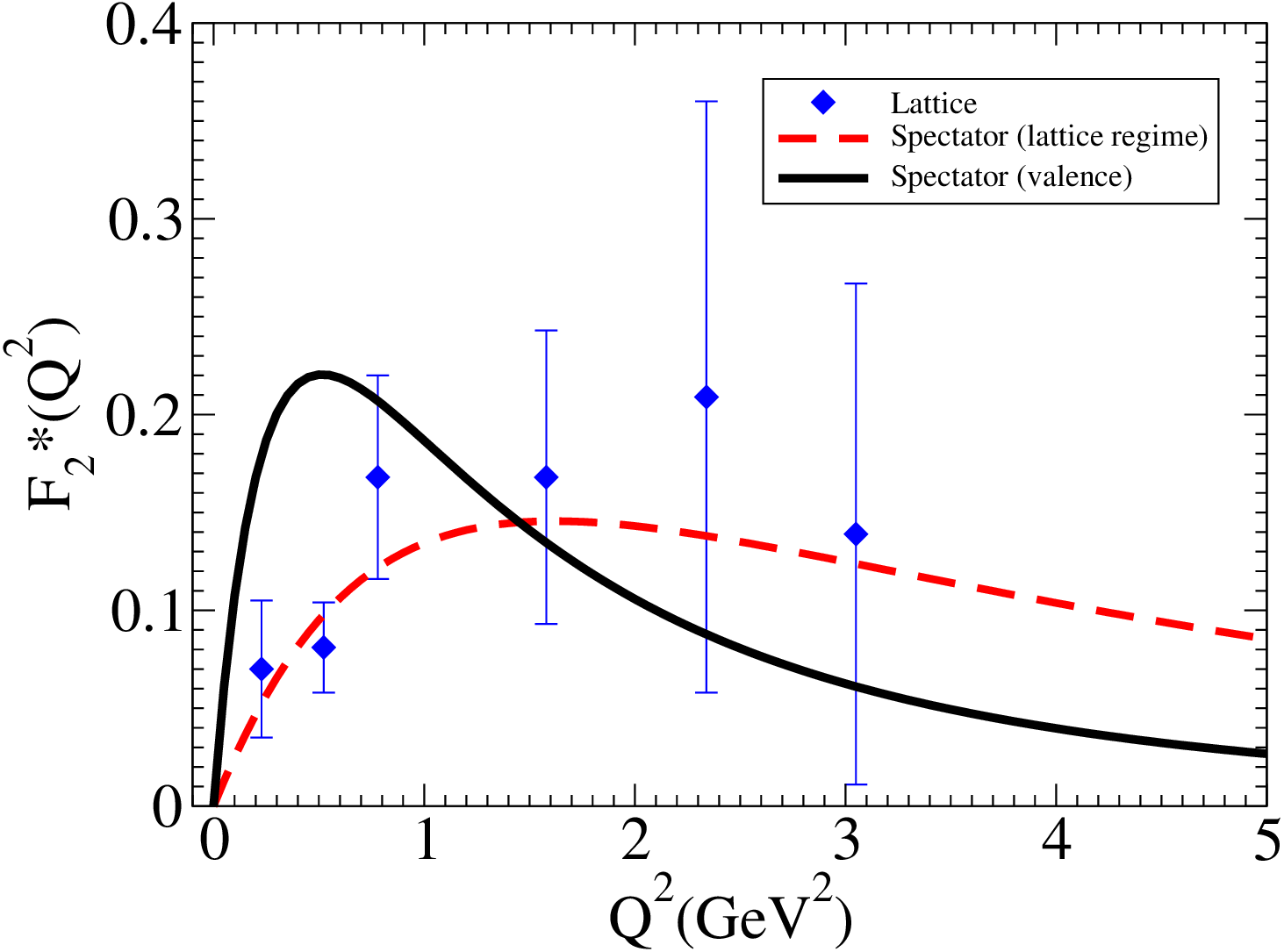}
}}
\caption{\footnotesize{
Valence quark contributions in the lattice regime (dashed lines)
corresponding to the pion mass $m_\pi =720$ MeV.
The lattice data with $m_\pi=720$ MeV (diamonds) are from Ref.~\cite{Lin09}.}
The solid lines are the valence quark contributions calculated with the physical
pion mass, $m_\pi =138$ MeV.
The mass values used corresponding to the lattice regime ($m_\pi=720$ MeV)
for the nucleon, Roper and $\rho$ meson are,
$M=1.48$ GeV, $M_R=2.53$ GeV, and $m_\rho=1.083$ GeV, respectively.
}
%\vspace{-1cm}
\label{figF1F2lat}
\end{figure*}

Next, we discuss the asymptotic
behavior for $F_1^\ast$ and $F_2^\ast$ as $Q^2 \to \infty$.
Perturbative QCD (pQCD)~\cite{Carlson,Sterman97}
predicts
$F_1^\ast(Q^2) \sim 1/Q^4$ and $F_2^\ast(Q^2) \sim 1/Q^6$
as $Q^2 \to \infty$,
apart from the $\log Q^2$ corrections.
The predictions of the present model from
Eqs.~(\ref{eqF1b})-(\ref{eqF2b}), are
consistent with these results:
$F_1^\ast(Q^2) \simeq 1.21\, {\cal I}$
and $F_2^\ast(Q^2) \simeq 13.1\, \sfrac{\cal I}{Q^2}$,
with ${\cal I}={\cal O}(\log Q^2)/Q^4$.
(See Appendix G of Ref.~\cite{NDelta} for details.)
For the helicity amplitudes,
our results are also consistent with the pQCD
predictions\footnote{
From Refs.~\cite{Carlson,Sterman97} one has
\be
G_+ \sim 1/Q^3, \hspace{.3cm} G_0 \sim 1/Q^4,
\ee
where $G_+ = A_{1/2}$ and $G_0 =\sfrac{Q}{|{\bf q}|}S_{1/2}$.
Note that the extra power in $G_0$ in contrast to $G_+$.
This takes account of changing the helicity one unit between
the initial and final states~\cite{Carlson,Sterman97}.
Thus, one gets, $A_{1/2} \sim 1/Q^3$ and $ S_{1/2} \sim 1/Q^3$.}:
$A_{1/2}, S_{1/2} \sim 1/Q^3$
\cite{Carlson,Sterman97},
again apart from the logarithmic corrections.
Note, however, that based on what observed for the
nucleon~\cite{Nucleon}, and especially for the
$\gamma N \to \Delta$ transition form factors~\cite{NDeltaD},
the predicted scaling behaviors may not be observed in such a small $Q^2$ region.
Also it is not realistic to expect that
our calibration of the quark current based on the
reactions in the regime $Q^2 < 6$ GeV$^2$,
can be naively extended to the high $Q^2$ region
where pQCD is valid (such as $Q^2 \sim 100-1000$ GeV$^2$).
%although the power law is consistent with pQCD.

\subsection{Comparing with the lattice results}
\label{complattice}

Extending the model to the lattice simulation regime,
we compare our results with the lattice QCD results.
Hereafter, the results obtained in the lattice regime
will be referred to as "lattice regime", which will be explained below.
One can expect that, the heavier the pion mass becomes
in the lattice simulations,
the closer our results become to the lattice one, since
the meson cloud effects become smaller.
A comparison is made in Fig.~\ref{figF1F2lat} for
the lattice data corresponding to $m_\pi=720$ MeV \cite{Lin09}.
To extend the model to the lattice regime,
we include an implicit dependence on the pion mass
for the hadron masses, following the procedure 
proposed in Refs.~\cite{Lattice,LatticeD}.
This is done by replacing the hadron masses in the model
by the respective lattice masses in the baryon wave function
and the quark electromagnetic current [Eq.~(\ref{eqjI})].
In the lattice regime (the dashed lines in Fig.~\ref{figF1F2lat})
we have used the nucleon and the Roper masses, $M= 1.48$ GeV and
$M_R=2.53$ GeV, respectively, corresponding to
the pion mass $m_\pi =720$ MeV \cite{Lin09}.
We have also used the value, 1.083 GeV for the $\rho$ meson mass,
according to the parameterization in Ref.~\cite{Leinweber01}.

One can see in Fig.~\ref{figF1F2lat} that the lattice regime results
are fairly consistent with the lattice data
for $F_2^\ast$.
As for $F_1^\ast$,
the lattice results overestimate our predictions for
$Q^2 > 1.5$ GeV$^2$, as well
as the covariant spectator model results
``Spectator (valence)''
[the same as those shown in Figs.~\ref{figF1F2} and~\ref{figF2}].
However, for $Q^2 < 0.6$ GeV$^2$, the agreement is excellent.
More lattice QCD simulation data with smaller pion masses
are desired to constrain better
the valence quark contributions in the low $Q^2$ region.
The agreement between the lattice regime
results and the lattice QCD data, supports that
the present estimate of the valence quark contributions
is reasonable.

\begin{figure*}[t]
\vspace{.5cm}
\centerline{
\mbox{
\includegraphics[width=3.2in]{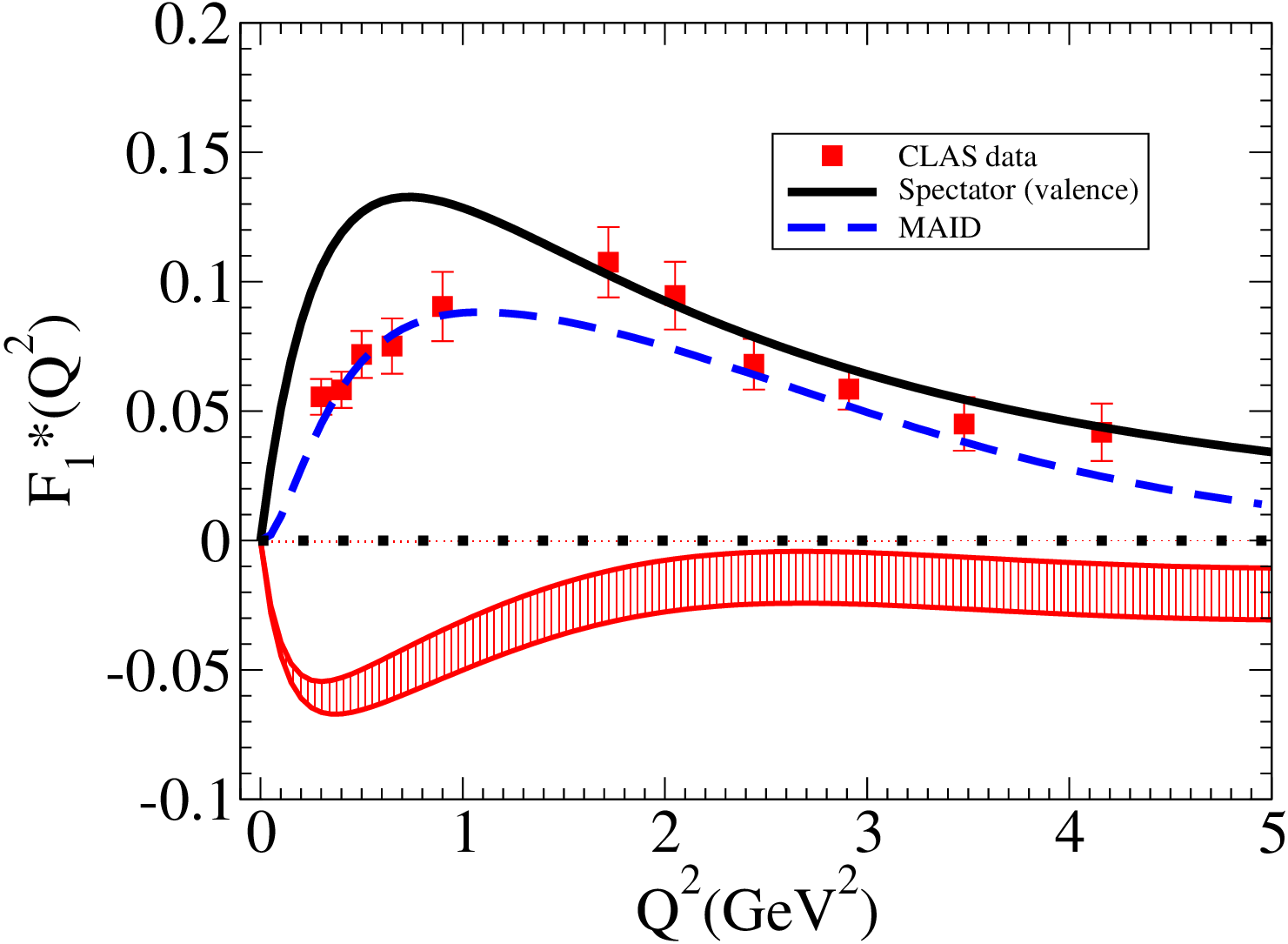}  \hspace{.1cm}
\includegraphics[width=3.2in]{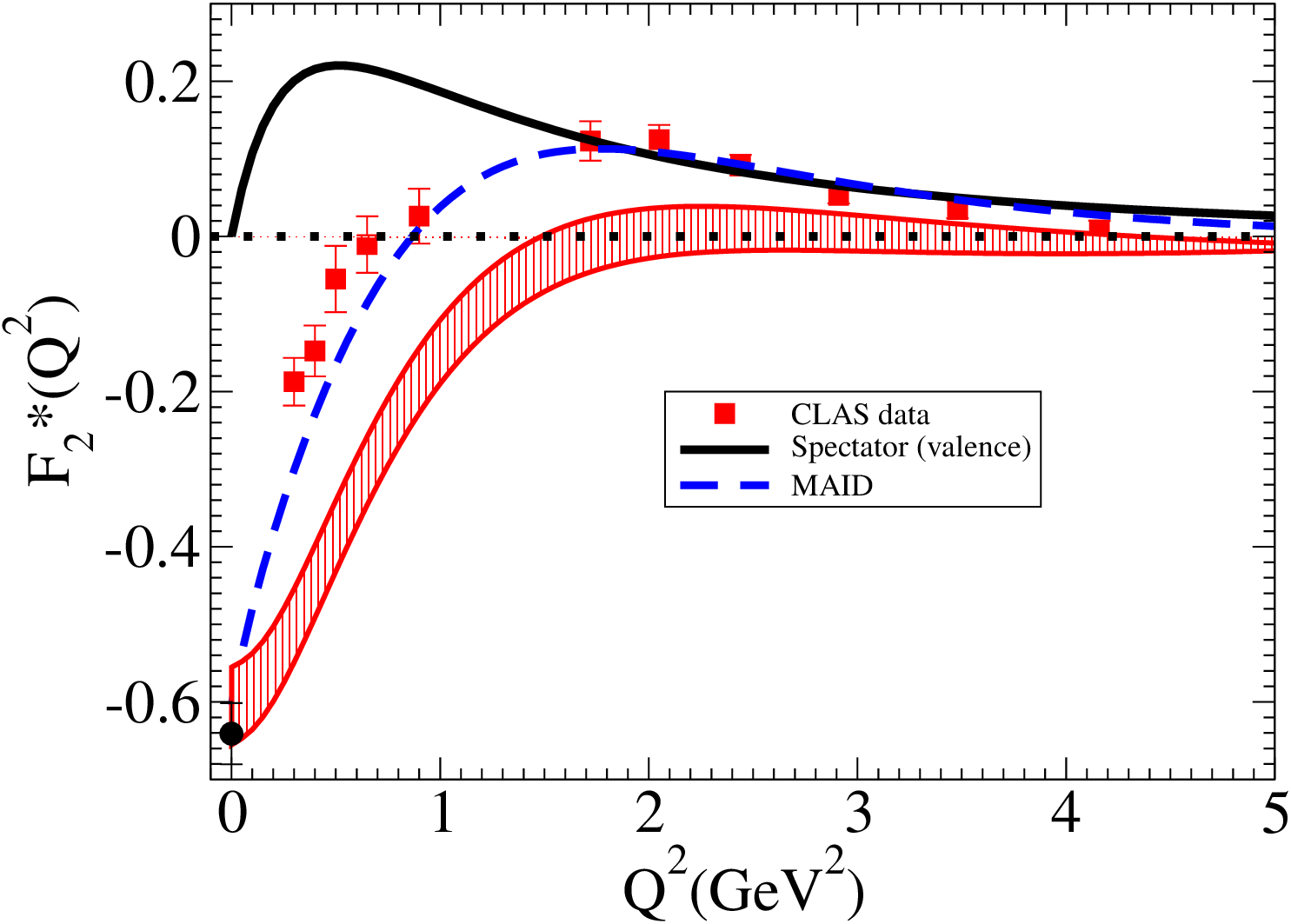}
}}
\caption{\footnotesize{
Meson cloud contributions (shaded areas) for $F_1^\ast$ (left panel)
and $F_2^\ast$ (right panel).
The bands are estimated by the upper limits of the CLAS error bars:
$\sigma(F_1^\ast) = 0.015(1-e^{-3Q^2})$
and $\sigma(F_2^\ast)= 0.05 -0.009 Q^2$, with one-standard deviation.
}}
%\vspace{-1cm}
\label{figMC}
\end{figure*}

\subsection{Estimating the meson cloud effects}

To estimate the meson cloud effects, we take
the valence quark contributions estimated in
Sec.~\ref{secPhysical} as a reference.
The valence quark contributions will be denoted by $F_i^{bare} (i=1,2)$.
We need to know a full contribution
to estimate the meson cloud contributions.
As a first step approach, we regard the MAID fit results~\cite{MAID},
which will be denoted by $F_i^{MAID}(Q^2) (i=1,2)$,
as the total contribution for the form factors.
Then, the meson cloud contributions $F_i^{mc}$,
may be estimated by
\be
F_i^{mc}(Q^2)= F_i^{MAID}(Q^2)-F_i^{bare}(Q^2), \hspace{1em}(i=1,2).
\ee
The results for $F_i^{mc}$ are presented in Fig.~\ref{figMC}.
To have an idea for the uncertainties in this
estimate we use an analytical form for the each form factor upper limit
of the CLAS data error bars [denoted by $\sigma(F_i^\ast) (i=1,2)$ below],
and calculate the bands with one-standard deviation.
Specifically, we use
$\sigma(F_1^\ast) \simeq 0.015(1-e^{-3Q^2})$
and  $\sigma(F_2^\ast) \simeq 0.05 -0.009 Q^2$.

First, we analyze $F_1^\ast$.
A situation for $F_1^\ast$ is a bit subtle.
The magnitude of the data is smaller than that of the $F_2^\ast$.
The upper limit for $|F_1^\ast|$ is around 0.1,
which is about half of that for $|F_2^\ast|$.
The meson cloud contributions are negative,
and they can be between 0.01 to 0.03 %(even)
in the high $Q^2$ region.
Probably, the contributions of 0.03 are overestimated, since
they are expected to become small in the high $Q^2$ region.
We note however, the meson cloud contributions here are estimated
by subtracting the model results from the MAID fit.
There are differences in the CLAS data and the MAID
fit in the high $Q^2$ region, since
the MAID analysis~\cite{CLAS} uses a different data set
from that of the CLAS~\cite{CLAS}.
The clarification between the differences in the two data sets,
and more precision data in this high $Q^2$ region
(less contaminations from meson cloud),
are desired to constrain better
the meson cloud contributions for $F_1^\ast$.

Next we discuss $F_2^\ast$.
Because of the larger magnitude of $F_2^\ast$, the contributions
from the meson cloud may be easier to estimate.
The meson cloud contributions for $F_2^\ast$
are negative as well as those for $F_1^\ast$.
As one can see in Fig.~\ref{figMC} the
meson cloud gives main contributions
in the low $Q^2$ region. (At $Q^2=0$, they are the only contributions.)
However, as expected, the meson cloud contributions
become small for $Q^2 > 1.5$ GeV$^2$,
and the valence quark contributions dominate.
In higher $Q^2$ region, the magnitude of
the meson cloud contributions are of order of the
error bars ($\sim 0.01$) of the data.

%%%%%%%%%%%%%%%%%%%
Overall, our estimate of the meson cloud contributions
qualitatively agree with those of
the recent works~\cite{JDiaz06,Li06,Golli09},
particularly for $F_1^\ast$.

In terms of the helicity amplitudes
the meson cloud contributions at $Q^2=0$
are estimated by $A_{1/2}^{mc}(0)=(-61.4\pm 4.9)\times 10^{-3}$ GeV$^{-1/2}$
and $S_{1/2}^{mc}(0)=(-40.2\pm 2.6)\times 10^{-3}$ GeV$^{-1/2}$.
Here the meson cloud is the only contribution for  $A_{1/2}(0)$.
As for $S_{1/2}(0)$, the meson cloud contributions
cancel significantly with those of the valence quark.
This feature can be understood better by observing Fig.~\ref{figA12S12}.

Thus, we predict negative contributions from the meson cloud for
$A_{1/2}$ and $S_{1/2}$ in the low $Q^2$ region.
This is consistent with the estimate
made in Ref.~\cite{Chen08}, where
meson cloud contributions are at least 70\% in $A_{1/2}(0)$
(100\% in our case), although in their case
the bare contributions differ in sign from our results.
A similar difference exists also for $S_{1/2}$,
in contradiction to the CLAS data in the high $Q^2$ region.
Also Ref.~\cite{Golli08a}
predicts significant negative
contributions from meson cloud for $S_{1/2}$ in the low $Q^2$ region,
implying large negative values for $S_{1/2}$
in that region, contrary to what is suggested by the CLAS data~\cite{CLAS}.
%and with the PDG result for $S_{1/2}(0)$ \cite{PDG}.
%This is contrary to what is suggested by the CLAS data~\cite{CLAS}.
%

We also attempt to compare our results
with those of the recent calculations based
on the dynamical meson-baryon
coupled-channel model of Sato and Lee~\cite{JDiaz08,JDiaz09}.
There, the bare contributions are very
small for $A_{1/2}(0)$~\cite{JDiaz08}.
In the region, $Q^2< 1$ GeV$^2$, the
absolute magnitudes of the meson cloud
contributions (their amplitudes are complex numbers)
are dominant in both amplitudes,
although the meson cloud contributions
decrease in the high $Q^2$ region
for $A_{1/2}$~\cite{JDiaz09}.
Unfortunately, their analysis~\cite{Suzuki09b} leads to the amplitudes different
from those of the CLAS.
Thus, it is expected that their meson cloud
contributions and ours differ quantitatively.

\section{Conclusions}
\label{secConclusions}

In this work we have studied the transition
form factors for $\gamma N \to P_{11}(1440)$
reaction using a covariant spectator quark model.
The model is fully relativistic,
and has no new adjustable parameters.
This is fulfilled by the
orthogonality condition between the nucleon and the Roper states.
The model can describe the reaction well in the high
$Q^2$ region (1.5 GeV$^2$ $< Q^2 < 4.0 $ GeV$^2$),
supporting the idea that the Roper is the first radial
excitation of the nucleon.
In the low $Q^2$ region ($Q^2 < 1.5$ GeV$^2$) the model predictions
deviate from the experimental data,
and this fact suggests that the meson cloud contributions
(missing in the model) are significant in this $Q^2$ region.

To check whether or not the description of our estimate
for the valence quark contributions is realistic,
we have extended the model to the lattice regime to compare
with the heavy-pion lattice QCD data.
Our results are in an excellent agreement with them in the low $Q^2$ region.
This fact supports that the valence quark degrees
of freedom is well described in our model.
Having a good confidence in the description of the valence
quark sector, we have estimated the meson cloud contributions
for the Dirac ($F_1^\ast$) and Pauli ($F_2^\ast$) type transition form factors.
A characteristic feature of the model,
$F_2^\ast(0)=0$, is a direct consequence of the
orthogonality condition between the nucleon and the Roper states.
However, this may not have significant consequences in the estimate of
the meson cloud contributions,
since they appear to be dominant at low $Q^2$, or
conversely, the valence quark contributions are expected to be
small at $Q^2=0$~\cite{Aznauryan07}.

In this work we have shown that the transition form factors
$F_1^\ast$ and $F_2^\ast$ for the $\gamma N \to P_{11}(1440)$ reaction
may be more suitable quantities to study than the helicity amplitudes
in order to disentangle the meson cloud contributions.
For example, $|F_1^\ast|$ is small for both in experimental data
and the valence quark model because of the cancellation between the $A_{1/2}$
and $S_{1/2}$ amplitudes. This implies also the meson cloud contributions are small.
As for $F_2^\ast$, it is mainly determined by $A_{1/2}(Q^2)$ in the very low $Q^2$ region,
which measures the meson cloud contributions in our approach (for $Q^2 \to 0$).

We admit that our estimate of the meson cloud contributions
has some limitations
due to the parameterization of the amplitudes
$A_{1/2}$ and $S_{1/2}$ based on the MAID analysis.
Furthermore, we have faced that there are some discrepancies between the
two data sets of the CLAS and MAID in the high $Q^2$ region.
More precision experimental data in this high $Q^2$ region
are desired to constrain better the meson cloud contributions.

For a possible improvement,
there is a room to add one extra free parameter to the model.
The parameter $\beta_1$ (fixed by the nucleon wave function) in the factor
$\sfrac{\beta_3 -\chi_R}{\beta_1+\chi_R}$ in the Roper wave function,
can be replaced by a new free parameter $\beta_4$ to introduce a new long range scale,
and may be adjusted by more precise, high $Q^2$ data.
Needless to say, more lattice QCD data are also useful
to constrain the valence quark contributions better.

We plan to extend the present model for higher mass resonance region,
like for instance, $S_{11}(1535)$.
In the past dynamical coupled-channel models
are very successful in the description
of the meson-baryon electro- and photoreaction,
involving  resonances such as $\Delta(1232)$, $P_{11}(1440)$
and  $S_{11}(1535)$.
However, such models require a parameterization in the interaction with
the quark core. In general, it is based on the baryon-meson phenomenology
and not based on the quark (and gluon) degrees of freedom.
The approach developed in this work is very
promising to study the valence quark contributions in
the meson-baryon systems.
It can be tested, or compared with
the bare parameters used in the dynamical coupled-channel models.
An independent test may be to compare
with the 'bare' contributions (no meson cloud)
determined by dynamical meson-baryon
coupled-channel models by fitting e.g.,
to the cross section data at each $Q^2$~\cite{JDiaz07,JDiaz09,Suzuki09b}.
This kind of an extraction of the
'bare' contributions was done for
the $\gamma N \to \Delta$ reaction~\cite{JDiaz07},
as well as other resonances including the Roper~\cite{JDiaz09,Suzuki09b}.
However, often the meson cloud contributions become larger than
the quark core contributions, and an estimate
of the bare contributions are sometimes difficult to
get as a smooth function of $Q^2$.
Another possible method is to use the lattice QCD data in the heavy-pion regime
to fix the valence quark contributions.
This method was successfully applied in Ref.~\cite{LatticeD}
for the nucleon to $\Delta$ electromagnetic transition.

\vspace{0.3cm}
\noindent
{\bf Acknowledgments:}

\vspace{0.2cm}

The authors thank B. Juli\'a-D{\'{i}}az, H. Kamano, M.~T.~Pe\~na,
A. Sibirtsev and A. Stadler for helpful discussions, and A.~W.~Thomas
for suggestions in the manuscript.
This work was support by Jefferson Science Associates,
LLC under U.S. DOE Contract No. DE-AC05-06OR23177,
and in part by the European Union
(HadronPhysics2 project ``Study of strongly interacting matter'').
G.~R.\ was also supported by the Portuguese Funda\c{c}\~ao para
a Ci\^encia e Tecnologia (FCT) under the Grant
No.~SFRH/BPD/26886/2006.
K.~T. acknowledges the CSSM (Adelaide, Australia) for hospitality,
where this work was completed.

\appendix

\section{Covariant spectator quark model}
\label{appSQM}

Below, we present some details of
the covariant spectator quark model.

\subsection{Wave functions}

In the covariant spectator quark model the S-state wave function
for the nucleon with the (initial) momentum $P_-$ is given by~\cite{Nucleon},
\be
\Psi_N(P_-,k)=
\frac{1}{\sqrt{2}}
\left[
\phi_I^0 u(P_-) -\phi_I^1
\varepsilon_{P_-}^\alpha U_\alpha(P_-)
\right] \psi_N(P_-,k),
\ee
where $k$ is the diquark momentum and $\phi_I^{0,1}$ is
the combination of the quark flavors
associated with the isospin 0 or 1 diquark.
The spin-1 polarization vector is represented by
$\varepsilon_{P_-}$ in a fixed-axis base~\cite{FixedAxis}, and
$u$ and $U^\alpha$ are the spinors related by~\cite{NDelta},
\be
U^\alpha (P)=
\frac{1}{\sqrt{3}} \gamma_5 \left(
\gamma^\alpha - \frac{P^\alpha}{M}\right)u(P).
\ee

The Roper wave function with the (final) momentum $P_+$ is defined by
\be
\Psi_R(P_+,k)=
\frac{1}{\sqrt{2}}
\left[
\phi_I^0 u_R(P_+) -\phi_I^1
\varepsilon_{P_+}^{\prime \alpha} U_{\alpha}^\prime(P_+)
\right] \psi_R(P_+,k),
\ee
where $u_R$ and $U_{\alpha}^\prime$ are
spin states associated with a spin 1/2 particle
and mass $M_R$, and $\varepsilon_{P_+}^{\prime \alpha}$ is
the polarization vector.

In both cases the scalar functions, $\psi_N(P_-,k)$ and $\psi_R(P_+,k)$, are
the functions of $(P_--k)^2$ and $(P_+-k)^2$, respectively.

\subsection{Quark form factors}

The quark current associated with Eqs.~(\ref{eqjI}) and~(\ref{eqjI2}) is expressed
in terms of the quark form factors $f_{i\pm} (i=1,2)$, inspired by
a vector meson dominance form:
\ba
& &
\hspace{-1.2cm}
f_{1\pm}(Q^2)=
\lambda + (1-\lambda)\frac{m_v^2}{m_v^2+Q^2}
+ c_\pm \frac{M_h^2Q^2}{(M_h^2+Q^2)^2}, \\
& &
\hspace{-1.2cm}
f_{2\pm} (Q^2)=
\kappa_\pm
\left\{
d_\pm
\frac{m_v^2}{m_v^2+Q^2}+
(1-d_\pm)\frac{M_h^2}{M_h^2+Q^2}
\right\}.
\ea
In the above, $\lambda$ defines
the quark charge in deep inelastic scattering,
$\kappa_\pm$ are the isoscalar and isovector
quark anomalous magnetic moments.
The mass $m_v$ ($M_h$) corresponds to
the light (heavy) vector meson,
and $c_\pm$, $d_\pm$ are the mixture coefficients.
In the present model we set $m_v =m_\rho$ ($\approx m_\omega$)
for the light vectorial meson
and $M_h = 2M$ (twice the nucleon mass)
to represent the short range physics.
The values of the parameters were previously fixed by
the nucleon elastic form factors~\cite{Nucleon},
and the values are presented in
Table~\ref{tabParam}. Note that the present model uses $d_+=d_-$.

\begin{table}[t]
\begin{center}
\begin{tabular}{c c c c c c c}
\hline
\hline
$\kappa_+$ & $\kappa_-$ & $c_+$ & $c_-$ & $d_+$ & $d_-$ & $\lambda$ \\
\hline
1.639 & 1.823 & 4.16 & 1.16 & $-0.686$ & $-0.686$ & 1.21 \\
\hline
\hline
\end{tabular}
\end{center}
\caption{Quark current parameters.}
\label{tabParam}
\end{table}

\subsection{Quark form factors and asymptotic expressions}

In the present model, the asymptotic expressions for $Q^2 \to \infty$,
associated with
the quark current of Eqs.~(\ref{eqjI2}) and~(\ref{eqjII}) are given by
\ba
& &
\hspace{-.5cm}
j_1 \simeq \frac{2}{3}\lambda,
\label{eqAj1}\\
& &
\hspace{-.5cm}
j_3 \simeq \frac{1}{6} (c_+-c_-) \frac{{\cal F}}{Q^2}, \\
& &
\hspace{-.5cm}
j_2 \simeq \frac{1}{6}\left( \kappa_+ + 3\kappa_-
\right) \frac{{\cal F}}{Q^2},
\label{eqj2} \\
& &
\hspace{-.5cm}
j_4 \simeq
\frac{1}{6} (\kappa_+- \kappa_-) \frac{\cal F }{Q^2},
\label{eqAj4}
\ea
with
\be
{\cal F}= d_+ m_v^2+ (1-d_+) M_h^2,
\ee
where, we have used the relation, $d_+=d_-$,
corresponding to the parameterization of the model II in Ref.~\cite{Nucleon}.
In particular, ${\cal F}=5.54$ GeV$^2$ should be noted.

Using Eqs.~(\ref{eqAj1})-(\ref{eqAj4}) we can derive
the expressions for $F_1^\ast$ and $F_2^\ast$ for $Q^2 \to \infty$.

\section{The orthogonality of the nucleon and the Roper states}
\label{appOrth}

In a static, nonrelativistic formalism the wave function
is a function of the particle's three-momentum, ${\bf k}$.
%We start considering the case where the
%initial and the final states have the same mass $M$.
In the limit where there is no momentum transfer,
the arguments in the initial ($\psi_i$) and final ($\psi_f$) wave functions
are the same.
This corresponds to ${\bf q}= {\bf k}-{\bf k}=0$, or
$Q^2= -{\bf q}^2=0$.
Then, the overlap integral between the orthogonal states, ${\cal I}$, is given by
\be
\int \frac{d^3 {\bf k}}{(2\pi)^3} \psi_f^*({\bf k})
\psi_i({\bf k})=0.
\label{eqOrth}
\ee

The equivalent relation for the covariant
spectator theory for Eq.~(\ref{eqOrth}) is
\be
\left. \int_k \psi_f^*(P_+,k) \psi_i (P_-,k)
\right|_{Q^2=0}=0,
\label{eqInt1}
\ee
where the subindex $Q^2=0$ indicates that
$Q^2=$ \mbox{$-(P_+ - P_-)^2=0$.}
%\mbox{$Q^2=-(P_+ - P_-)^2=0$.}
The simplest case is in the initial (or final)
baryon's rest frame. For the equal mass case,
$P_+=P_-=(M,0,0,0)$.

Next, consider the inelastic case
with the masses of the initial and final states, $M$ and $M_R$, respectively.
The generalization of the condition Eq.~(\ref{eqInt1})
would correspond to
$P_-=(M,0,0,0)$ and $P_+=(M_R,0,0,0)$, with
$q=P_+ - P_- =\left(\sfrac{M_R^2-M^2-Q^2}{2M_R},0,0,|{\bf q}|\right)$
when ${\bf q}={\bf 0}$.
($|{\bf q}|$ is given by Eq.~(\ref{eqq2}).)
This gives also
\be
Q^2\equiv Q^{\ast 2}= -(M_R-M)^2,
\ee
which will be denoted by the pseudo-threshold point,
the point where both initial and final state are at rest.
This situation is, however, unphysical for $M_R \ne M$
because $Q^2 <0$.

The generalization of Eq.~(\ref{eqInt1}) for
unequal mass case would be,
\be
\left. \int_k \psi_f^*(P_+,k) \psi_i (P_-,k)
\right|_{Q^2=Q^{\ast 2}}=0.
\label{eqInt2}
\ee
Since the physical reaction are restricted to $Q^2 \ge 0$,
we need to redefine the orthogonality condition in the unequal mass case.
To do so we impose the condition,
\be
\int_k \psi_f^*(\bar P_+,k) \psi_i (\bar P_-,k)=0,
\label{eqInt3}
\ee
where the four-momenta $\bar P_+$ and $\bar P_-$
are respectively defined by,
\ba
& &
\bar P_-=\left( \frac{M_R^2+M^2}{2M_R},0,0,-\frac{M_R^2-M^2}{2M_R}\right),
\nonumber \\
& &
\bar P_+=(M_R,0,0,0), %\nonumber
\ea
which correspond to $Q^2=0$, but ${\bf q}^2 \ne 0$.

The use of Eq.~(\ref{eqInt3}) may be regarded as an approximation,
or the simplest relativistic extension for the orthogonality condition.
The exact treatment needs to impose
the overlap integral to vanish at the pseudo-threshold
as in Eq.~(\ref{eqInt2}).
However, it would require an extension of the wave functions
to the unphysical region, and beyond a scope of the present study.
Instead, we use Eq.~(\ref{eqInt3}).
Both prescriptions should give similar results when $(M_R-M)^2$ is small enough.


\vspace{.15cm}
\begin{references}

%\begin{itemize}
\bibitem{Clement06}
  H.~Clement {\it et al.},
  %``Evidence for a 'narrow' Roper resonance: The breathing mode of the
  %nucleon,''
  arXiv:nucl-ex/0612015.
  %%CITATION = NUCL-EX/0612015;%%
\bibitem{Sarantsev08}
  A.~V.~Sarantsev {\it et al.},
  %``New results on the Roper resonance and the $P_{11}$ partial wave,''
  Phys.\ Lett.\  B {\bf 659}, 94 (2008)
  [arXiv:0707.3591 [hep-ph]].
  %%CITATION = PHLTA,B659,94;%%
\bibitem{Suzuki09}
  N.~Suzuki, B.~Julia-Diaz, H.~Kamano, T.~S.~Lee, A.~Matsuyama and T.~Sato,
  %``Structure and dynamical evolution of low lying nucleon resonances,''
  arXiv:0909.1356 [nucl-th].
  %%CITATION = ARXIV:0909.1356;%%
\bibitem{Aznauryan07}
  I.~G.~Aznauryan,
  %``Electroexcitation of the Roper resonance in the relativistic quark
  %models,''
  Phys.\ Rev.\  C {\bf 76}, 025212 (2007)
  [arXiv:nucl-th/0701012].
  %%CITATION = PHRVA,C76,025212;%%
\bibitem{CLAS}
  I.~G.~Aznauryan {\it et al.}  [CLAS Collaboration],
  %``Electroexcitation of nucleon resonances from CLAS data on single pion
  %electroproduction,''
  Phys.\ Rev.\  C {\bf 80}, 055203 (2009)
  [arXiv:0909.2349 [nucl-ex]].
  %%CITATION = PHRVA,C80,055203;%%
\bibitem{MAID}
  D.~Drechsel, S.~S.~Kamalov and L.~Tiator,
  %``Unitary Isobar Model - MAID2007,''
  Eur.\ Phys.\ J.\  A {\bf 34}, 69 (2007)
  [arXiv:0710.0306 [nucl-th]].
  %%CITATION = EPHJA,A34,69;%%
\bibitem{Tiator09}
  L.~Tiator and M.~Vanderhaeghen,
  %``Empirical transverse charge densities in the nucleon-to-P11(1440)
  %transition,''
  Phys.\ Lett.\  B {\bf 672}, 344 (2009)
  [arXiv:0811.2285 [hep-ph]].
  %%CITATION = PHLTA,B672,344;%%
\bibitem{Aznauryan08}
  I.~G.~Aznauryan, V.~D.~Burkert and T.~S.~Lee,
  %``On the definitions of the gamma*N -> N* helicity amplitudes,''
  arXiv:0810.0997 [nucl-th].
  %%CITATION = ARXIV:0810.0997;%%
\bibitem{Burkert08}
  V.~D.~Burkert,
  %``Electromagnetic Transition Form Factors of Nucleon Resonances,''
  AIP Conf.\ Proc.\  {\bf 1056}, 348 (2008)
  [arXiv:0808.2326 [nucl-ex]].
  %%CITATION = APCPC,1056,348;%%
\bibitem{JDiaz08}
  B.~Julia-Diaz, T.~S.~H.~Lee, A.~Matsuyama, T.~Sato and L.~C.~Smith,
  %``Dynamical Coupled-Channels Effects on Pion Photoproduction,''
  Phys.\ Rev.\  C {\bf 77}, 045205 (2008)
  [arXiv:0712.2283 [nucl-th]].
  %%CITATION = PHRVA,C77,045205;%
\bibitem{Nucleon}
  F.~Gross, G.~Ramalho and M.~T.~Pe\~na,
  %``A pure S-wave covariant model for the nucleon,''
  Phys.\ Rev.\  C {\bf 77}, 015202 (2008)
  [arXiv:nucl-th/0606029].
\bibitem{Li92}
  Z.~P.~Li,
  %``Photoproduction Signatures Of Hybrid Baryons: An Application Of The Quark
  %Model With Gluonic Degrees Of Freedom,''
  Phys.\ Rev.\  D {\bf 44}, 2841 (1991);
  %%CITATION = PHRVA,D44,2841;%%
%\bibitem{Li92}     % 116 citations
  Z.~Li, V.~Burkert and Z.~Li,
  %``Electroproduction of the Roper resonance as a hybrid state,''
  Phys.\ Rev.\  D {\bf 46}, 70 (1992).
  %%CITATION = PHRVA,D46,70;%%
\bibitem{Warns90} % 43 citation
  M.~Warns, W.~Pfeil and H.~Rollnik,
  %``Helicity And Isospin Asymmetries In The Electroproduction Of Nucleon
  %Resonances,''
  Phys.\ Rev.\  D {\bf 42}, 2215 (1990);
  %%CITATION = PHRVA,D42,2215;%%
% Study of Relativistic effects
%\bibitem{Warns90} % 36 citations
  M.~Warns, H.~Schroder, W.~Pfeil and H.~Rollnik,
  %``Calculations Of Electromagnetic Nucleon Form-Factors And Electroexcitation
  %Amplitudes Of Isobars,''
  Z.\ Phys.\  C {\bf 45}, 627 (1990).
  %%CITATION = ZEPYA,C45,627;%%
\bibitem{Capstick95}
  S.~Capstick and B.~D.~Keister,
  %``Baryon current matrix elements in a light front framework,''
  Phys.\ Rev.\  D {\bf 51}, 3598 (1995)
  [arXiv:nucl-th/9411016].
  %%CITATION = PHRVA,D51,3598;%%
  %$[${\bf Relativistic and non-relativistic expressions}$].$
\bibitem{Cardarelli98}  % ***** reference
  F.~Cardarelli, E.~Pace, G.~Salme and S.~Simula,
  %``Electroproduction of the Roper resonance and the constituent quark
  %model,''
  Phys.\ Lett.\  B {\bf 397}, 13 (1997)
  [arXiv:nucl-th/9609047].
  %%CITATION = PHLTA,B397,13;%%
  %$[${\bf Ligh-front CQM}$].$
\bibitem{Giannini01}   % 28 citations
  M.~M.~Giannini, E.~Santopinto and A.~Vassallo,
  %``Hypercentral constituent quark model and isospin dependence,''
  Eur.\ Phys.\ J.\  A {\bf 12}, 447 (2001)
  [arXiv:nucl-th/0111073].
  %%CITATION = EPHJA,A12,447;%%
\bibitem{JDiaz04}
  B.~Julia-Diaz, D.~O.~Riska and F.~Coester,
  %``Baryon Form Factors of Relativistic Constituent-Quark Models,''
  Phys.\ Rev.\  C {\bf 69}, 035212 (2004)
  [Erratum-ibid.\  C {\bf 75}, 069902 (2007)]
  [arXiv:hep-ph/0312169].
  %%CITATION = PHRVA,C69,035212;%%
\bibitem{Weber90}
  H.~J.~Weber,
  %``ELECTROEXCITATION OF THE N* (1440) IN THE RELATIVISTIC CONSTITUENT QUARK
  %MODEL,''
  Phys.\ Rev.\  C {\bf 41}, 2783 (1990).
  %%CITATION = PHRVA,C41,2783;%%
\bibitem{Cano98}
  F.~Cano and P.~Gonzalez,
  %``A consistent explanation of the Roper phenomenology,''
  Phys.\ Lett.\  B {\bf 431}, 270 (1998)
  [arXiv:nucl-th/9804071].
  %%CITATION = PHLTA,B431,270;%%
\bibitem{Li06}     %  30% in Roper
  Q.~B.~Li and D.~O.~Riska,
  %``The role of q anti-q components in the N(1440) resonance,''
  Phys.\ Rev.\  C {\bf 74}, 015202 (2006)
  [arXiv:nucl-th/0605076].
  %%CITATION = PHRVA,C74,015202;%%
\bibitem{Santopinto08}
  E.~Santopinto and R.~Bijker,
  %``Quark-antiquark effects in baryons,''
  Few Body Syst.\  {\bf 44}, 95 (2008).
  %%CITATION = FBSYE,44,95;%%
\bibitem{Dillig04}
  M.~Dillig and M.~Schott,
  %``Mesonic content of the nucleon and the Roper resonance,''
  Phys.\ Rev.\  C {\bf 75}, 067001 (2007)
  [Erratum-ibid.\  C {\bf 76}, 019903 (2007)]
  [arXiv:nucl-th/0405063].
  %%CITATION = PHRVA,C75,067001;%%
\bibitem{JDiaz06}
  B.~Julia-Diaz and D.~O.~Riska,
  %``The role of q q q q anti-q components in the nucleon and the N(1440)
  %resonance,''
  Nucl.\ Phys.\  A {\bf 780}, 175 (2006)
  [arXiv:nucl-th/0609064].
  %%CITATION = NUPHA,A780,175;%%
\bibitem{Krehl00}   % 122 citations
  O.~Krehl, C.~Hanhart, C.~Krewald and J.~Speth,
  %``What is the structure of the Roper resonance?,''
  Phys.\ Rev.\  C {\bf 62}, 025207 (2000)
  [arXiv:nucl-th/9911080].
  %%CITATION = PHRVA,C62,025207;%%
\bibitem{JDiaz07b}   % 32 citations
  B.~Julia-Diaz, T.~S.~Lee, A.~Matsuyama and T.~Sato,
  %``Dynamical Coupled-Channel Model of $\pi N$ Scattering in the W $\leq$ 2 GeV
  %Nucleon Resonance Region,''
  Phys.\ Rev.\  C {\bf 76}, 065201 (2007)
  [arXiv:0704.1615 [nucl-th]].
  %%CITATION = PHRVA,C76,065201;%%
\bibitem{JDiaz07}
  B.~Julia-Diaz, T.~S.~H.~Lee, T.~Sato and L.~C.~Smith,
  %``Extraction and interpretation of gamma N --> Delta form factors within a
  %dynamical model,''
  Phys.\ Rev.\  C {\bf 75}, 015205 (2007)
  [arXiv:nucl-th/0611033].
  %%CITATION = PHRVA,C75,015205;%%
\bibitem{JDiaz09}
  B.~Julia-Diaz, H.~Kamano, T.~S.~H.~Lee, A.~Matsuyama, T.~Sato and N.~Suzuki,
  %``Dynamical coupled-channels analysis of p(e,e'pi)N reactions,''
  Phys.\ Rev.\  C {\bf 80}, 025207 (2009)
  [arXiv:0904.1918 [nucl-th]].
\bibitem{Suzuki09b}
  N.~Suzuki, T.~Sato and T.~S.~Lee,
  %``Extraction of Transition Form Factors for Nucleon Resonances within a
  %Coupled-Channels Model,''
  arXiv:0910.1742 [nucl-th].
  %%CITATION = ARXIV:0910.1742;%%
\bibitem{Golli09}
  B.~Golli, S.~Sirca and M.~Fiolhais,
  %``Pion electro-production in the Roper region in chiral quark models,''
  Eur.\ Phys.\ J.\  A {\bf 42}, 185 (2009)
  [arXiv:0906.2066 [nucl-th]].
  %%CITATION = EPHJA,A42,185;%%
\bibitem{Golli08a}
  B.~Golli and S.~Sirca,
  %``Excitation of the Roper resonance,''
  Few Body Syst.\  {\bf 44}, 157 (2008).
  %%CITATION = FBSYE,44,157;%%
\bibitem{Julich}
%\bibitem{Schneider06}
  S.~Schneider, S.~Krewald and U.~G.~Meissner,
  %``The reaction pi N --> pi pi N in a meson-exchange approach,''
  Eur.\ Phys.\ J.\  A {\bf 28}, 107 (2006)
  [arXiv:nucl-th/0603040].
  %%CITATION = EPHJA,A28,107;%%
\bibitem{Kamalov01}
  S.~S.~Kamalov, S.~N.~Yang, D.~Drechsel, O.~Hanstein and L.~Tiator,
  %``gamma* N --> Delta transition form factors: A new analysis of the JLab
  %data on p(e,e' p)pi0 at Q**2 = 2.8-(GeV/c)**2 and 4.0-(GeV/c)**2,''
  Phys.\ Rev.\  C {\bf 64}, 032201(R) (2001)
  [arXiv:nucl-th/0006068].
  %%CITATION = PHRVA,C64,032201;%%
\bibitem{Burkert04}
  V.~D.~Burkert and T.~S.~H.~Lee,
  %``Electromagnetic meson production in the nucleon resonance region,''
  Int.\ J.\ Mod.\ Phys.\  E {\bf 13}, 1035 (2004)
  [arXiv:nucl-ex/0407020].
  %%CITATION = IMPAE,E13,1035;%%
\bibitem{Bermuth90}  % 50 citations
  K.~Bermuth, D.~Drechsel, L.~Tiator and J.~B.~Seaborn,
  %``PHOTOPRODUCTION OF DELTA AND ROPER RESONANCES IN THE CLOUDY BAG MODEL,''
  Phys.\ Rev.\  D {\bf 37}, 89 (1988).
  %%CITATION = PHRVA,D37,89;%%
\bibitem{Dong99}
  Y.~B.~Dong, K.~Shimizu, A.~Faessler and A.~J.~Buchmann,
  %``Description of the electroproduction amplitudes of the Roper resonance in a
  %relativistic quark model,''
  Phys.\ Rev.\  C {\bf 60}, 035203 (1999).
  %%CITATION = PHRVA,C60,035203;%%
\bibitem{Chen08}
  D.~Y.~Chen and Y.~B.~Dong,
  %``Electroproduction of Roper resonance in a meson cloud model,''
  Commun.\ Theor.\ Phys.\  {\bf 50}, 142 (2008).
  %%CITATION = CTPMD,50,142;%%
%\cite{Lin:2008qv}
\bibitem{Lin09}
  H.~W.~Lin, S.~D.~Cohen, R.~G.~Edwards and D.~G.~Richards,
  %``First Lattice Study of the $N$-$P_{11}(1440)$ Transition Form Factors,''
  Phys.\ Rev.\  D {\bf 78}, 114508 (2008)
  [arXiv:0803.3020 [hep-lat]].
  %%CITATION = PHRVA,D78,114508;%%
\bibitem{Gross}
  F.~Gross,
  %``Three-Dimensional Covariant Integral Equations For Low-Energy Systems,''
  Phys.\ Rev.\  {\bf 186}, 1448 (1969);
  %%CITATION = PHRVA,186,1448;
  %\bibitem{Gross92}
  F.~Gross, J.~W.~Van Orden and K.~Holinde,
  %``Relativistic One Boson Exchange Model For The Nucleon-Nucleon
  %Interaction,''
  Phys.\ Rev.\ C {\bf 45}, 2094 (1992).
  %%CITATION = PHRVA,C45,2094;%%
\bibitem{NDelta}
  G.~Ramalho, M.~T.~Pe\~na and F.~Gross,
  %``A Covariant model for the nucleon and the $\Delta$,''
  Eur.\ Phys.\ J.\  A {\bf 36}, 329 (2008)
  [arXiv:0803.3034 [hep-ph]].
  %%CITATION = EPHJA,A36,329;%%
\bibitem{NDeltaD}
  G.~Ramalho, M.~T.~Pe\~na and F.~Gross,
  %``D-state effects in the electromagnetic N-Delta transition,''
  Phys.\ Rev.\  D {\bf 78}, 114017 (2008)
  [arXiv:0810.4126 [hep-ph]].
  %%CITATION = PHRVA,D78,114017;%%
\bibitem{DeltaFF0}
  G.~Ramalho and M.~T.~Pe\~na,
  %``Electromagnetic form factors of the Delta in a S-wave approach,''
  J.\ Phys.\ G {\bf 36}, 085004 (2009)
  [arXiv:0807.2922 [hep-ph]];
  %%CITATION = JPHGB,G36,085004;%%
%\bibitem{DeltaQuad}
  G.~Ramalho, M.~T.~Pe\~na and F.~Gross,
  %``Electric quadrupole and magnetic octupole moments of the Delta,''
  Phys.\ Lett.\  B {\bf 678}, 355 (2009)
  [arXiv:0902.4212 [hep-ph]].
  %%CITATION = PHLTA,B678,355;%%
\bibitem{DeltaFF}
  G.~Ramalho, M.~T.~Pe\~na and F.~Gross,
  %``Electromagnetic form factors of the Delta with D-waves,''
  arXiv:1002.4170 [hep-ph].
  %%CITATION = ARXIV:1002.4170;%%
\bibitem{Omega}
  G.~Ramalho, K.~Tsushima and F.~Gross,
  %``A relativistic quark model for the Omega- electromagnetic form factors,''
  Phys.\ Rev.\  D {\bf 80}, 033004 (2009)
  [arXiv:0907.1060 [hep-ph]].
  %%CITATION = PHRVA,D80,033004;%%
\bibitem{Octet}
  F.~Gross, G.~Ramalho and K.~Tsushima,
  %``Using baryon octet magnetic moments and masses to fix the pion cloud
  %contribution,''
  arXiv:0910.2171 [hep-ph].
  %%CITATION = ARXIV:0910.2171;%%
%\cite{Wang:2007iw}
\bibitem{Wang:2007iw}
  P.~Wang, D.~B.~Leinweber, A.~W.~Thomas and R.~D.~Young,
  %``Chiral extrapolation of nucleon magnetic form factors,''
  Phys.\ Rev.\  D {\bf 75}, 073012 (2007)
  [arXiv:hep-ph/0701082].
  %%CITATION = PHRVA,D75,073012;%%
\bibitem{PDG}
  C.~Amsler {\it et al.}  [Particle Data Group],
  %``Review of particle physics,''
  Phys.\ Lett.\  B {\bf 667}, 1 (2008).
  %%CITATION = PHLTA,B667,1;
\bibitem{Carlson}
  C.~E.~Carlson,
  %``Electromagnetic N - Delta Transition At High Q**2,''
  Phys.\ Rev.\  D {\bf 34}, 2704 (1986);
  %%CITATION = PHRVA,D34,2704;%%
\bibitem{Sterman97}
  G.~Sterman and P.~Stoler,
  %``Hadronic form factors and perturbative QCD,''
  Ann.\ Rev.\ Nucl.\ Part.\ Sci.\  {\bf 47}, 193 (1997)
  [arXiv:hep-ph/9708370].
  %%CITATION = ARNUA,47,193;%%
\bibitem{Lattice}
  G.~Ramalho and M.~T.~Pe\~na,
  %``Nucleon and gamma N -> Delta lattice form factors in a constituent quark
  %model,''
  J.\ Phys.\ G {\bf 36}, 115011 (2009)
  [arXiv:0812.0187 [hep-ph]].
  %%CITATION = JPHGB,G36,115011;%%
\bibitem{LatticeD}
  G.~Ramalho and M.~T.~Pe\~na,
  %``Valence quark contribution for the gamma N -> Delta quadrupole transition
  %extracted from lattice QCD,''
  Phys.\ Rev.\  D {\bf 80}, 013008 (2009)
  [arXiv:0901.4310 [hep-ph]].
  %%CITATION = PHRVA,D80,013008;%%
\bibitem{Leinweber01}
  D.~B.~Leinweber, A.~W.~Thomas, K.~Tsushima and S.~V.~Wright,
  %``Chiral behaviour of the rho meson in lattice QCD,''
  Phys.\ Rev.\  D {\bf 64}, 094502 (2001)
  [arXiv:hep-lat/0104013].
  %%CITATION = PHRVA,D64,094502;%%
\bibitem{FixedAxis}
  F.~Gross, G.~Ramalho and M.~T.~Pe\~na,
  %``Fixed-axis polarization states: covariance and comparisons,''
  Phys.\ Rev.\  C {\bf 77}, 035203 (2008).
  %%CITATION = PHRVA,C77,035203;%%





\end{references}
\end{document}